\documentclass[12pt,english]{article}
\usepackage[english]{babel}
\usepackage{amsmath,amssymb,amscd}
\usepackage{amsfonts}
\usepackage{indentfirst}
\usepackage[dvips]{graphicx}
\usepackage{youngtab}

\makeatletter
\renewcommand\section{\@startsection {section}{1}{\z@}%
                                   {-5.5ex \@plus -1ex \@minus -.2ex}
                                   {2.3ex \@plus.2ex}%
                                   {\normalfont\large\bfseries}}
\renewcommand\subsection{\@startsection{subsection}{2}{\z@}%
                                     {-3.25ex\@plus -1ex \@minus -.2ex}%
                                     {1.5ex \@plus .2ex}%
                                     {\normalfont\bfseries}}

\newcommand\subsub[1]{
\bigskip

\noindent{\underline{\it #1}}
\smallskip}

\makeatother

\newcommand{\bea}{\begin{eqnarray}}
\newcommand{\eea}{\end{eqnarray}}
\newcommand{\be}{\begin{equation}}
\newcommand{\ee}{\end{equation}}

\newcommand{\Z}{{\mathbb Z}}
\newcommand{\R}{{\mathbb R}}
\newcommand{\C}{{\mathbb C}}

\renewcommand{\P}{{\mathbb P}}

\newcommand{\Q}{{\mathbb Q}}

\def\Tr{{\rm Tr}}

\def\a{\alpha}

\def\G{\Gamma}

\def\la{\lambda}

\def\t{\tau}

\def\w{\omega}

\def\lieg{{\mathfrak g}}
   
\newcommand{\cB}{{\cal B }}            
\newcommand{\cM}{{\cal M }}            
\newcommand{\cF}{{\cal F }}

\newcommand{\cO}{{\cal O }}            
            
\newcommand{\cA}{{\cal A }}            

\newcommand{\cN}{{\cal N }}            
\newcommand{\cY}{{\cal Y }}    
\newcommand{\cD}{{\cal D }} 
\newcommand{\cV}{{\cal V }}    
\newcommand{\E}{{\cal E }}   
\newcommand{\B}{{\cal B}} 
\renewcommand{\d}{{\partial}}

\def\wt{\widetilde}

\begin{document}

\sloppy

\baselineskip=18pt

\begin{flushright}
\begin{tabular}{l}
HUTP-07/A006\\
ITFA-2007-44\\
IFT-UW/2007-9\\
 [.3in]
\end{tabular}
\end{flushright}

\begin{center}
{\Large \sc Supersymmetric Gauge Theories,\\[4mm]
 Intersecting Branes and Free Fermions}\\[16mm]

Robbert Dijkgraaf,$\!{}^{1,2}$ Lotte Hollands,$\!{}^1$ 
Piotr Su{\l}kowski$\!{}^{\, 1,3,4}$ and Cumrun Vafa$^5$\\[6mm]

\emph{$^1$\!\! Institute for Theoretical Physics, and \,$^2$\!\! KdV Institute
  for Mathematics,} \\ 
\emph{University of Amsterdam, Valckenierstraat
  65, 1018 XE Amsterdam, The Netherlands} \\[4mm]
\emph{$^3$\!\! Institute for Theoretical Physics, Warsaw University,} \\
\emph{and \,$^4$\!\! So{\l}tan Institute for Nuclear Studies,} \\
\emph{ul. Ho\.za 69, PL-00-681 Warsaw, Poland} \\[4mm]
\emph{$^5$\!\! \it
Jefferson Physical Laboratory, Harvard University, Cambridge, MA
02138, USA}
\end{center}

\vspace{1cm}

\centerline{\bf Abstract}
\smallskip
\begin{quote}
We show that various holomorphic quantities in supersymmetric gauge
theories can be conveniently computed by configurations of D4-branes
and D6-branes. These D-branes intersect along a Riemann surface that
is described by a holomorphic curve in a complex surface. The
resulting I-brane carries two-dimensional chiral fermions on its
world-volume. This system can be mapped directly to the topological
string on a large class of non-compact Calabi-Yau manifolds. Inclusion
of the string coupling constant corresponds to turning on a constant
$B$-field on the complex surface, which makes this space
non-commutative. Including all string loop corrections the free
fermion theory is elegantly formulated in terms of holonomic
$D$-modules that replace the classical holomorphic curve in the
quantum case. 

\end{quote}

\newpage


\section{Introduction}

Substantial progress has been made in understanding four-dimensional
supersymmetric gauge theories in terms of elegant exact solutions. A
constant factor in all these solutions has been the relation to
two-dimensional geometry and conformal field theory. Many quantities
in gauge theory have turned out to be expressible in terms of an
effective Riemann surface or complex curve $\Sigma$ and a particular
quantum field theory living on this curve. An ubiquitous role in all
this is played by free fermion systems.

Perhaps the first example has been Montonen-Olive S-duality in $\cN=4$
supersymmetric gauge theories \cite{montonen-olive}. These field
theories are invariant under $SL(2,\Z)$ transformations of the
complexified gauge coupling $\tau$.  This includes in particular the
strong-weak coupling S-duality $\tau \to -1/\tau$.

These S-dualities are closely related to the modular invariance of a
CFT on a two-torus. Indeed, in certain cases the partition function on
a four-manifold $M$ has been shown to exactly reproduce the character
of a two-dimensional conformal field theory \cite{vafa-witten}. This
connection was first shown in the beautiful mathematical work of
Nakajima, who showed that in the case of ALE singularities there
exists an action of an affine Kac-Moody algebra on the cohomology of
the instanton moduli space \cite{nakajima}. Many of the CFT's that
arise in this fashion are closely related to free fermion systems or
equivalently chiral bosons. A well-known example is that of a $K3$
manifold, which gives the partition function of the heterotic
string. These relations between four-dimensional and two-dimensional
systems are most naturally understood by considering six-dimensional
theories on the space $M \times T^2$, a connection that we will make
good use of in this paper.

A second example is the celebrated Seiberg-Witten
\cite{seiberg-witten} solution of $\cN=2$ gauge theories, which
involves a spectral curve $\Sigma$ of general genus. Many properties
of the gauge theory are captured by the geometry of this curve. For
example, BPS masses are related to the periods of a particular
meromorphic one-form on $\Sigma$ and the holomorphic gauge coupling
matrix $\tau_{IJ}(t)$, that appears in the low-energy $U(1)^N$ abelian
gauge theory Lagrangian as
$$
\int d^4x \ \tau_{IJ} F_+^I \wedge F_+^J,
$$ 
can be identified with the period matrix of $\Sigma$ as
a function of the moduli $t$.

In $\cN=1$ a closely related structure arises because certain
holomorphic quantities, such as the superpotential and gauge
couplings, can be computed exactly by sums over planar diagrams
\cite{dijkgraaf-vafa}. The corresponding large $N$ matrix model can be
solved in terms of an effective geometry that again in many cases
takes the form of a Riemann surface $\Sigma$ endowed with a particular
meromorphic one-form.

These relations extend beyond classical field theory on $\Sigma$. For
example, in both $\cN=2$ and $\cN=1$ theories one can compute the
effective action that results from coupling the field theory to a
four-dimensional curved background metric \cite{uplane,kmt,dst}. In
these cases the coefficient $\cF_1$ for the coupling $R_+ \wedge R_+$
takes a particularly nice form: it can be expressed as
$$
\cF_1(t) = - {1\over 2} \log \det \Delta_\Sigma
$$
where $\Delta_\Sigma$ is the (chiral) Laplacian on the curve
$\Sigma$. So, using the boson-fermion correspondence, the
gravitational coupling in four dimensions is captured by a quantum
field theory of free fermions on $\Sigma$.

F-terms in supersymmetric gauge theories are closely related to
topological string theories, since these gauge theories can be
geometrically engineered by considering specific decoupling limits of
type II string compactifications on well-chosen Calabi-Yau manifolds
\cite{geo-engin,kmv-mirror,h-i-v}. In many cases the relevant
non-compact Calabi-Yau manifold takes the form
$$
uv+ P(x,y) =0,
$$ 
where $P(x,y)=0$ defines the corresponding curve $\Sigma$ embedded in
$\C^2$ or $\C^* \times \C^*$. In this fashion also a role can be given
for the higher loop amplitudes $\cF_g$ of the topological string. In
\cite{adkmv} it has been shown that also these higher genus terms can 
be computed using free fermions on $\Sigma$, but now the fermions
should be given a ``quantum character'' closely related to a
quantization of the coordinates $x,y$. We will elucidate this
phenomenon at the end of this paper using the formalism of $\cal
D$-modules.

Also Nekrasov has shown that by working equivariantly with respect to
the $U(1)$ action on $\R^4$, the higher $\cF_g$
terms make an appearance in $\cN=2$ theories \cite{nekrasov,nekrasov-okounkov}. From this point of view the corresponding integrable sytems are solved 
naturally by means of free fermions too.

In this paper we shed some new light on the ubiquity of these
free fermion systems. Our strategy will be to map the supersymmetric
gauge theory to a system of intersecting D-branes. One immediate
advantage of this reformulation is that it makes the relation between 
$\cN=4$ gauge theories on ALE spaces and two-dimensional CFT
transparent. 

These intersecting branes turn out to also provide a natural setting for
understanding the higher genus corrections $\cF_g$ in terms of
non-commutative geometry. In fact, one of our conclusions will be that
in the full quantum theory these fermions should not be considered
as local fields, but as sections of a $\cD$-module.

\newpage

\vspace*{2mm}
\begin{figure}[htb*]
\setlength{\unitlength}{0.16cm}
\centering
\begin{picture}(100,106)

\put(0,80){\framebox(32,12){$\begin{array}{c} \textrm{\bf IIA} \\
\underbrace{\R^3\times \big(\Sigma \subset\cB\big)\times\R^2\times
  S^1}_{\textrm{I-brane on}\ \Sigma\ =\ N\, \textrm{D4}\,\cap\, k\, \textrm{D6}} \end{array}$}}
\put(68,80){\framebox(32,12){$\begin{array}{c} \textrm{\bf IIA} \\
\underbrace{TN_k\times (\Gamma}_{N\,\textrm{D4/NS5}} \subset \cB_3) \times \R^2 \times S^1 \end{array}$}}

\put(34,100){\framebox(32,12){$\begin{array}{c} \textrm{\bf M-theory} \\ 
\underbrace{ TN_k\times\big(\Sigma}_{N\,\textrm{M5}} \subset\cB\big)\times\R^2\times S^1 \end{array}$}}
\put(34,80){\framebox(32,12){$\begin{array}{c} \textrm{\bf IIA} \\
\underbrace{TN_k\times \big(\Sigma}_{N\,\textrm{NS5}} \subset \cB \big) \times \R^2\end{array}$}}
\put(34,60){\framebox(32,12){$\begin{array}{c} \textrm{\bf IIB} \\
TN_k\times X_{\Sigma}  \end{array}$}}
\put(34,40){\framebox(32,12){$\begin{array}{c} \textrm{\bf IIA} \\
TN_k\times {\wt X}_{\textrm{toric}}  \end{array}$}}
\put(34,20){\framebox(32,12){$\begin{array}{c} \textrm{\bf M-theory} \\
TN_k\times S^1 \times {\wt X}_{\textrm{toric}}  \end{array}$}}
\put(34,0){\framebox(32,12){$\begin{array}{c} \textrm{\bf IIA} \\
\R^3 \times \underbrace{S^1 \times {\wt X}_{\textrm{toric}}}_{k\,\textrm{D6}}  \end{array}$}}

\put(34,100){\vector(-2,-1){15}} \put(1,95){\small reduce on $S^1_{TN}$}
\put(66,100){\vector(2,-1){15}} \put(79,95){\small reduce on $S^1_\Sigma$}
\put(40,100){\vector(0,-1){7.5}} \put(42,95){\small reduce on extra $S^1$}
\put(40,80){\vector(0,-1){7.5}} \put(42,75){\small T-dualize }
\put(40,60){\vector(0,-1){7.5}} \put(42,55){\small mirror symmetry}
\put(40,40){\vector(0,-1){7.5}} \put(42,35){\small lift to M-theory}
\put(40,20){\vector(0,-1){7.5}} \put(42,15){\small reduce on $S^1_{TN}$}

\end{picture}
\end{figure}
\begin{center}
{\small \bf Fig.\ 1: \it Web of dualities considered in the paper.}
\end{center}


Our considerations will be based on a chain of dualities, relating
various string and M-theory compactifications with branes and
fluxes. For convenience we present these dualities, whose precise
details will be explained subsequently, in figure 1.

The plan of this paper is as follows: in section 2 we discuss $\cN=4$
supersymmetric gauge theories on ALE and Taub-NUT manifolds. We show
how results of Nakajima are naturally reproduced by mapping the gauge
theory, realized via D4-branes in IIA string theory, to an
intersecting brane system on a $T^2$. A crucial role is played by the
level-rank duality. Section 3 generalizes this approach to arbitrary
Riemann surfaces. In this way we make contact with $\cN=2$ and $\cN=1$
gauge theories on one hand, and Calabi-Yau compactifications of type
II superstrings on the other. Finally, in section 4 we show that
higher genus amplitudes of the topological string in these backgrounds
can be captured by turning on a $B$-field, making the geometry
non-commutative. The appropriate mathematical formalism to understand
the free fermions turns out to be the theory of $\cD$-modules. We
illustrate this formalism of a conformal field theory of $\cD$-modules
with some concrete examples in (space-time) genus 0, 1, and 2.


\section{$\mathcal{N}=4$ theories on ALE spaces}

Let us start by considering a $U(N)$ $\cN=4$ supersymmetric gauge
theory on a, possibly non-compact, four-manifold $M$. We will take $M$
to be a hyper-K\"ahler manifold.  In this case the ordinary $\cN=4$
supersymmetric gauge theory is equivalent to a topologically twisted
one (as some supercharges are preserved on hyper-K\"ahler manifolds).
Including the topological couplings $\theta$ and $v \in H^2(M,\Z)$,
the action is given by
$$ 
S = - \int {i\over 4 \pi} \tau\, \Tr(F_+\wedge F_+) + v \wedge \Tr\,F_+
+ \hbox{\it c.c.},
$$
where the complexified gauge coupling $\tau$ is given by
\be
\label{tau}
\tau = {\theta\over 2\pi} + {4\pi i \over g^2}.
\ee
Here complex conjugation not only changes $\tau$ and $v$ into their
anti-holomorphic conjugates, but also maps the self-dual part $F_+$ of
the field strength to the anti-self-dual part $F_-$. In the
topologically twisted theory only instantons contribute\footnote{In our
  choice of twisting (and conventions) these are given by self-dual
  solutions with $F_-=0$.}  and the partition function of the $U(N)$
gauge theory becomes a holomorphic function of the coupling $\tau$ (up
to possible holomorphic anomalies). The $v$-dependence is entirely
captured by the $U(1)$ factor and is in general given in terms of
Siegel theta-functions of signature $(b^2_+,b^2_-)$
$$
\sum_{p \in H^2(M,\Z)} e^{i\pi (\tau p_+^2 - \bar\tau p_-^2)} e^{2\pi
  i(v \cdot p_+ - \bar{v} \cdot p_-)}.
$$
Here $p$ and $v$ are elements of $H^2(M, \Z)$, so that $v \cdot p$ (and 
likewise $p^2$) refers to the intersection product $\int_M v \wedge p$.
Clearly this contribution only becomes holomorphic in the case that
$b^2_-=0$.

Because of S-duality the partition function $Z(v,\t)$ of this gauge
theory is expected to be given by a Jacobi form determined by the
geometry $M$ \cite{vafa-witten}. That is, it should have the following 
transformation properties:
\bea
Z\left({v \over c\tau + d},{a\tau +b \over c\tau +d} \right) 
 = (c\tau + d)^w e^{2\pi i \kappa c v^2/(c\tau +d)}  Z(v,\tau), 
&& 
\left[\begin{array}{cc}
a & b \\
c & d \\
\end{array}\right] \in SL(2,\Z). \nonumber \\
Z(v + n\tau + m, \tau) = e^{-2\pi i \kappa (n^2\tau + 2n \cdot v)}
Z(v,\tau),
&&
n,m \in H^2(M,\Z) \cong \Z^{b_2}. \nonumber
\eea
The weight of the Jacobi-form $w=-\chi(M)$ is given by minus the
(regularized) Euler number of $M$ and its index is $\kappa=N$.

Using the localization to instantons, the partition function has a
Fourier expansion of the form
$$
Z(v,\tau) = \sum_{m\in H^2(M),n\geq 0} d(m,n)\, y^m q^{n-c/24},
$$
where $y = e^{2\pi i v}$, $q = e^{2\pi i \tau}$ and $c=N \chi(M)$.
The coefficients $d(n,m)$ are roughly computed as the Euler number of
the moduli space of $U(N)$ instantons on $M$ with total instanton
numbers $c_1=m$ and $ch_2=n$. 

In general the coefficients $d(m,n)$ are believed to be integers,
because they have a direct interpretation as computing BPS invariants
in a five-dimensional gauge theory. If we consider a D4-brane wrapping
the five-manifold $M \times S^1$, then we can compute $d(m,n)$ as the
index\footnote{Here and in the subsequent sections we assume that the two
  fermion zero modes associated to the center of mass movements of the
  D4-brane have been absorbed.}
$$
d(m,n) = \Tr (-1)^F \in \Z,
$$ 
in the subsector of field configurations on $M$ of given instanton
numbers $m,n$. Here we interpret the $S^1$ as Euclidean time.

From the string theory point of view the modular invariance of $Z$ is
explained naturally by lifting the D4-brane to M-theory, where it
becomes an M5-brane on the product manifold
$$
M \times T^2.
$$ 
The world-volume theory of the M5-brane is (in the low-energy limit)
the rather mysterious six-dimensional $U(N)$ conformal field theory
with $(0,2)$ supersymmetry. The complexified gauge coupling $\tau$ can
now be interpreted as the modulus of the elliptic curve $T^2$, while
the Wilson loops of the 3-form potential $C_3$ along this curve are
related to the couplings $v$, as we explain in more detail in section
\ref{ssec-D4D6}.  With this interpretation the action of modular group
$SL(2,\Z)$ on $v$ and $\tau$ is the obvious geometric one.

\subsection{Gauge theory on ALE spaces}

We now want to turn to a very concrete case, where we take $M$ to be a
non-compact ALE space. Such a manifold can be written as the (possibly
resolved) orbifold
$$
M_\G \to \C^2/\G,
$$ 
with $\G$ a finite subgroup of $SU(2)$. These Kleinian singularities
have an ADE classification. For the $A_{k-1}$ singularity, that we
will mostly restrict to in this paper, $\G$ is given by the cyclic group
$\Z_k$.

We do however have to address the fact that the four-manifold $M_\G$
is non-compact, so that we have to fix boundary conditions for the
gauge field. The boundary at infinity is given by the Lens space
$S^3/\G$ and here the $U(N)$ gauge field should approach a flat
connection. Up to gauge equivalence this flat connection is labeled
by an $N$-dimensional representation of the quotient group $\Gamma$,
that is, an element
$$
\rho \in {\rm Hom}\left(\G,U(N)\right).
$$
If $\rho_i$ label the irreducible representations of $\Gamma$, then 
$\rho$ can be decomposed as
$$
\rho= \bigoplus_i N_i \rho_i,
$$
where the multiplicities $N_i$ are non-negative integers satisfying
the restriction
$$
\sum_i N_i d_i = N, \qquad d_i = \dim \rho_i.
$$
Now the famous McKay correspondence \cite{mckay,mckay-slodowy}
$$
\G \ \leftrightarrow\ \widehat{\lieg},
$$ 
relates the finite subgroups $\G \subset SU(2)$ with corresponding
simple Lie algebras $\lieg$ of ADE type, or more properly their
affine extensions $\widehat{\lieg}$. We will denote the compact Lie group
corresponding to the Lie algebra $\lieg$ as
$G$. In the McKay correspondence the irreducible representations
$\rho_i$ of the finite group $\G$ are related to the nodes of the
extended Dynkin diagram of the affine algebra $\widehat{\lieg}$. The
dimensions $d_i$ of these irreps can then be identified with the dual
Dynkin indices.

Through the McKay correspondence each $N$-dimensional representation
$\rho$ of $\G$ determines an integrable highest-weight representation
of $\widehat{\lieg}_N$ at level $N$. We will denote this
(infinite-dimensional) Lie algebra representation as $V_\rho$. In
particular for $\Gamma=\Z_k$, which is the case that we will mostly
concentrate on, flat connections on $S^3/\Z_k$ thus get identified
with integrable representations of $\widehat{su}(k)_N$. In this particular
case all Dynkin indices satisfy $d_i=1$.

With $\rho$ labeling the boundary conditions of the gauge field at
infinity, we will get a vector-valued partition function
$Z_\rho(v,\tau)$. Formally the $U(N)$ gauge theory partition function on the
ALE manifold again has an expansion
$$
Z_\rho(v,\tau) = \sum_{n,m} d(m,n) y^m q^{h_\rho+n-c/24},
$$ 
where $c=N k$ with $k$ the regularized Euler number of the $A_{k-1}$
manifold \cite{vafa-witten}.  The usual instanton numbers given by the
second Chern class $n=ch_2$ in the exponent are now shifted by a
rational number $h_{\rho}$, which is related to the Chern-Simons
invariant of the flat connection $\rho$. As we explain in section
\ref{ssec-mckay}, $h_{\rho}$ gets mapped to the conformal dimension of
the corresponding integrable weight in the affine Lie algebra
$\widehat{\lieg}$ related to $\Gamma$ by the McKay correspondence.
S-duality will act non-trivially on the boundary conditions $\rho$,
and therefore $Z_\rho(v,\tau)$ will be a vector-valued Jacobi form
\cite{vafa-witten}.

For these ALE spaces the instanton computations can be explicitly
performed, because there exists a generalized ADHM construction in
which the instanton moduli space is represented as a quiver
variety. The physical intuition underlying this formalism has been
justified by the beautiful mathematical work of Nakajima
\cite{nakajima,nakajima2}, who has proven that on the middle
dimensional cohomology of the instanton moduli space one can actually
realize the action of the affine Kac-Moody algebra $\widehat{\mathfrak
  g}_N$ in terms of geometric operations. In fact, this work leads to
the identification
$$
Z_{\rho}(v,\tau) = \Tr_{\strut V_\rho}\!\left(y^{J_0} q^{L_0-c/24} \right) =
\chi_\rho(v,\tau),
$$ 
with $V_\rho$ the integrable highest-weight representation of
$\widehat{\lieg}_N$ and $\chi_\rho$ its affine character. Here $c$ is
the appropriate central charge of the corresponding WZW model. A
remarkable fact is that, in the case of a $U(N)$ gauge theory on a
$\Z_k$ singularity, we find an action of $\widehat{su}(k)_N$ and not
of the gauge group $SU(N)$. This is a important example of the
familiar level-rank duality of affine Lie algebras.

Now, interestingly, Frenkel has suggested
\cite{frenkel} that, if one works equivariantly for the action of the
gauge group $SU(N)$ at infinity (we ignore the $U(1)$ part for the
moment), there would similarly be an action of the $\widehat{su}(N)_k$
affine Lie algebra.  Physically this means ``ungauging'' the
$SU(N)$ at infinity.  In other words, we consider making the $SU(N)$ into a
global symmetry instead of a gauge symmetry at the boundary.  This
suggestion has recently been confirmed in \cite{licata}. So, depending
on how we deal with the theory at infinity, there are reasons to
expect both affine symmetry structures to appear and have a combined
action of the Lie algebra
$$
\widehat{su}(N)_K \times \widehat{su}(k)_N.
$$ 
We will now turn to a dual string theory realization, where this
structure indeed becomes transparent.

\subsection{The Taub-NUT geometry}

In order to study this gauge system within string theory, we use a
trick that proved to be very effectively in relating 4d and 5d black
holes \cite{gsy,gsy2,recount-dyons,dvv,bdf} and is in line with
the duality between ALE spaces and 5-brane geometries \cite{ns5-ale}.
We will replace the local $A_{k-1}$
singularity with a Taub-NUT geometry. This can be best understood as
a $S^1$ compactification of the singularity. The $TN_k$ geometry is a
hyper-K\"ahler manifold with metric \cite{Ruback,Sen-TN},
$$
ds^2_{TN} = R^2 \left[ {1\over V}(d\chi
+ \alpha)^2 + V d{\vec x}^2 \right],
$$
with $\chi \in S^1$ (with period $4\pi$) and ${\vec x}\in \R^3$. Here
the function $V$ and 1-form $\a$ are determined as
$$ 
V(\vec{x}) = {1 + \sum_{a=1}^k {1\over |\vec{x} - {\vec x}_a|}}, \qquad
d \alpha = *_3 \,\, dV.
$$ 
The Taub-NUT manifold can be thought as a (singular) circle fibration
$$
\begin{matrix}
S^1 & \to & TN_k \\
& &  \downarrow     \\[2mm]
& & \R^3
\end{matrix}
$$
where the size of the $S^1$ shrinks at the points
$\vec{x}_1,\ldots,\vec{x}_k \in \R^3$. These positions are the hyperk\"ahler
moduli of the space. The total manifold is however perfectly
smooth. At infinity it approximates the cylinder $\R^3 \times S^1$
where the $S^1$ has fixed radius $R$, but is non-trivially fibered
over the $S^2$ at infinity as a monopole bundle of charge (first
Chern class) $k$
$$
\int_{S^2} d\alpha = 2\pi k. 
$$
In the core, where we can ignore the constant $1$ that appears in the
expression for the potential $V(\vec{x})$, the Taub-NUT geometry can be
approximated by the (resolved) $A_{k-1}$ singularity.

The manifold $TN_k$ has non-trivial 2-cycles $C_{a,b} \cong S^2$ that
are fibered over the line segments joining the locations $\vec{x}_a$ and
$\vec{x}_b$ in $\R^3$. Only $k-1$ of these cycles are homologically
independent. As a basis we can pick the cycles
$$
C_a:=C_{a,a+1}, \qquad a=1,\ldots,k-1.
$$
The intersection matrix of these 2-cycles gives the Cartan matrix of
$A_{k-1}$.

From a dual perspective, there are $k$ independent normalizable
harmonic 2-forms $\w_a$ on $TN_k$, that can be chosen to be localized
around the centers or NUTs $\vec{x}_a$. With 
$$
V_a={1\over |\vec{x}-\vec{x}_a|}, \qquad
d\alpha_a=*dV_a,
$$
they are given as
$$
\w_a = d \eta_a,\qquad \eta_a = \alpha_a-{V_a \over V}(d\chi +\alpha).
$$
Furthermore, these 2-forms satisfy
$$
\int_{TN} \w_a \wedge \w_b = 16 \pi^2 \delta_{ab},
$$
and are dual to the cycles $C_{a,b}$
$$
\int_{C_{a,b}} \w_c = 4 \pi (\delta_{ac} - \delta_{bc}).
$$
A special role is played by the sum of these harmonic 2-forms
\be \w = \sum_a \w_a.
\label{omega}
\ee
This is the unique normalizable harmonic 2-form that is invariant under
the tri-holomorphic $U(1)$ isometry of $TN$. The form $\w$ has zero
pairings with all the cycles $C_{ab}$.  In the ``decompactification
limit'', where $TN_k$ gets replaced by $A_{k-1}$, the linear
combination $\w$ becomes non-normalizable, while the $k-1$ two-forms
orthogonal to it survive.  

We will make convenient use of the following elegant interpretation of
the two-form $\w$.  Consider the $U(1)$ action on the $TN_k$ manifold that rotates
the $S^1$ fiber. It is generated by a Killing vector field $\xi$. Let
$\eta$ be the corresponding dual one-form given as
$
\eta_\mu = g_{\mu\nu} \xi^\nu,
$
where we used the $TN$-metric to convert the vector field to a
one-form. Up to an overall rescaling this gives
\be
\label{eta}
\eta = {1\over V}(d\chi + \alpha).
\ee
In terms of this one-form, $\w$ is given by
$\w = d \eta.$

\subsection{String theory realization}  \label{ssec-realization}

Our strategy in this paper will be that, since we consider the twisted
partition function of the topological field theory, the answer will be
formally independent of the radius $R$ of the Taub-NUT geometry. So we
can take both the limit $R \to \infty$, where we recover the result
for the ALE space $\C^2/\Z_k$, and the limit $R \to 0$, where the
problem becomes essentially three-dimensional.

Now, there are some subtleties with this argument, since a priori the
partition function of the gauge theory on the $TN$ manifold is
\emph{not} identical to that of the ALE space. In particular there are
new topological configurations of the gauge field that can
contribute. These can be thought of as monopoles going around the
$S^1$ at infinity. We will come back to this subtle point later.

In type IIA string theory, the partition function of the $\cN=4$ SYM
theory on the $TN_k$ manifold can be obtained by considering a
compactification of the form
$$
\hbox{(IIA)}\quad TN \times S^1 \times \R^5,
$$
and wrapping $N$ D4-branes on $TN \times S^1$. This is a special case of 
the situation presented in the box on the right-hand side in fig.~1,
with $\Gamma=S^1$, $\cB_3=S^1\times\R^2$, and $S^1$ decompactified.
In the decoupling limit the partition function of this set of D-branes will
reproduce the Vafa-Witten partition function on $TN_k$. This partition
function can be also written as an index
$$
Z(v,\tau) = \Tr \left((-1)^F
e^{-\beta H}  e^{i n \theta}e^{2\pi i m v} \right) 
$$
where $\beta=2\pi R_9$ is the circumference of the ``9th dimension''
$S^1$, and $m=c_1$, $n=ch_2$ are the Chern characters of the gauge
bundle on the $TN_k$ space. Here we can think of the theta angle $\theta$
as the Wilson loop for the graviphoton field $C_1$ along the
$S^1$. Similarly $v$ is the Wilson loop for $C_3$. The gauge coupling
of the 4d gauge theory is now identified as
$$
{1\over g^2} = { \beta \over g_s \ell_s}.
$$ 
Because only BPS configurations contribute in this index, again only
the holomorphic combination $\tau$ (\ref{tau}) will appear.

We can now further lift this configuration to M-theory
with an additional $S^1$ of size  $R_{11} = g_s l_s$, where we
obtain the compactification
$$
\hbox{(M)}\quad TN \times T^2 \times \R^5,
$$ 
now with $N$ M5-branes wrapping the six-manifold $TN_k \times T^2$. 
This corresponds to the top box in fig.~1, with $\Sigma=T^2$.
As we remarked earlier, after this lift the coupling constant $\tau$ is
interpreted as the geometric modulus of the elliptic curve $T^2$.
In particular its imaginary part is given by the ratio $R_9/R_{11}$.
Dimensionally reducing the six-dimensional $U(N)$ theory on the
M5-brane world-volume over the Taub-NUT space gives a two-dimensional
$(0,8)$ superconformal field theory, in which the gauge theory
partition function is computed as the elliptic genus
$$
Z = \Tr \left((-1)^F y^{J_0} q^{L_0-c/24} \right).
$$

In order to further analyze this system we switch to yet another
duality frame by compactifying back to Type IIA theory, but now
along the $S^1$ fiber in the Taub-NUT geometry. This is the familiar
9-11 exchange. In this fashion we end up with a IIA compactification
on
$$
\hbox{(IIA)}\quad \R^3 \times T^2 \times \R^5,
$$
with $N$ D4-branes wrapping $\R^3 \times T^2$. However, because the
circle fibration of the $TN$ space has singular points, we have to
include D6-branes as well. In fact, there will be $k$ D6-branes that wrap $T^2
\times \R^5$ and are localized at the points
$\vec{x}_1,\ldots,\vec{x}_k$ in the $\R^3$. 
This situation is represented in the box on the left-hand side in fig.~1.

Summarizing, we get a system of $N$ D4-branes and $k$ D6-branes intersecting
along the $T^2$. This intersection locus is called the I-brane and it
is depicted in fig.~2.  We will now study this I-brane
system in greater detail.

\begin{figure}[t]
\begin{center}   \label{fig1}
\includegraphics[width=8cm]{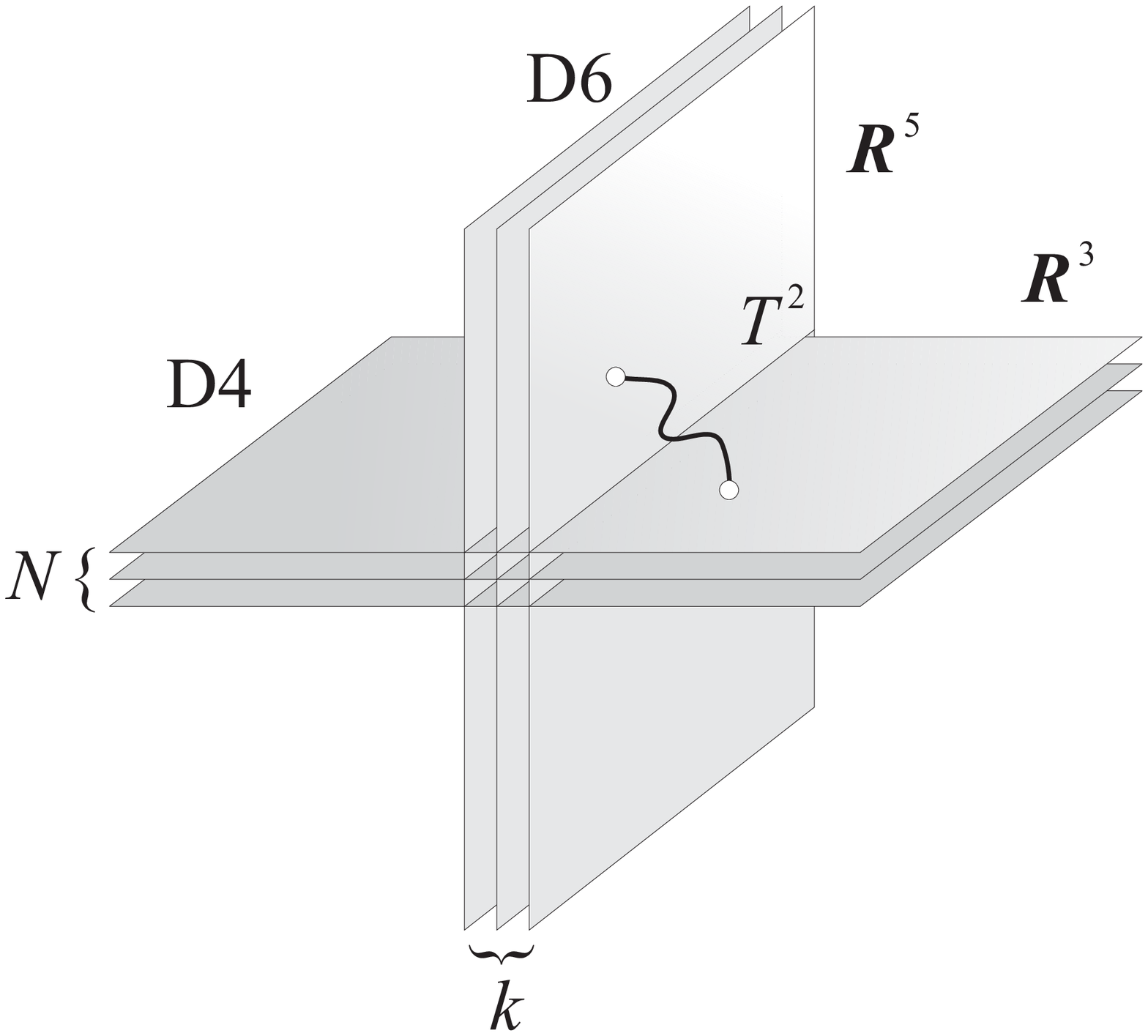}\\[5mm]
\parbox{12cm}{\small \bf Fig.\ 2: \it Configuration of
  intersecting D4 and D6-branes with one of the 4-6 open strings that
  gives rise to a chiral fermion localized on the I-brane.}
\end{center} 
\end{figure}

\subsection{The D4-D6 system and free fermions}  \label{ssec-D4D6}

A collection of D4-branes and D6-branes that intersect along two
(flat) dimensions is a supersymmetric configuration. One way to see
this is that after some T-dualities, it can be related to a D0-D8 or
D1-D9 system. The supersymmetry in this case is of type $(0,8)$. The
massless modes of the 4-6 open strings stretching between the D4 and
D6 branes reside entirely in the Ramond sector. All modes in the NS
sector are massive. These massless modes are well-known to be chiral
fermions on the two-dimensional I-brane \cite{Ibrane,qm80,Ibrane46}. If we have
$N$ D4-branes and $k$ D6-branes, the chiral fermions
$$
\psi_{i,\overline a}(z),\,\, \psi^\dagger_{\overline\imath, a}(z), \qquad
i=1,\ldots,N,\ a=1,\ldots,k
$$ 
transform in the bifundamental representations $(N,\overline{k})$ and
$(\overline{N},k)$ of $U(N) \times U(k)$. Since we are computing an
index, we can take the $\alpha' \to 0$ limit, in which all massive
modes decouple. In this limit we are just 
left with the chiral fermions. Their action is necessarily free and given
by 
$$ I = \int d^2z\ \psi^\dagger \bar\partial_{A + \wt A}\psi,
$$ 
where $A$ and $\wt A$ are the restrictions to the I-brane $T^2$ of the
$U(N)$ and $U(k)$ gauge fields, that live on the worldvolumes of the
D4-branes and the D6-branes respectively. (Here we have absorbed the overall  
coupling constant). 

Under the two $U(1)$'s the fermions have charge $(+1,-1)$. Therefore
the overall (diagonal) $U(1)$ decouples and the fermions
effectively  couple to the gauge group
$$
U(1) \times SU(N) \times SU(k),
$$ 
where the remaining $U(1)$ is the anti-diagonal. At this point we ignore  
certain discrete identifications under the $\Z_N$ and $\Z_k$ centers,
that we will return to later.

\subsub{Zero modes}

A special role is played by the zero-modes of the D-brane gauge
fields. In the supersymmetric configuration we can have both a
non-trivial flat $U(N)$ and $U(k)$ gauge field turned on along the
$T^2$.  We will denote these moduli as $u_i$ and $v_a$
respectively.
The partition function of the chiral fermions on the I-brane will be a
function $Z(u,v,\tau)$ of both the flat connections $u,v$ and the
modulus $\tau$. It will transform as a (generalized) Jacobi-form under
the action of $SL(2,\Z)$ on the two-torus.

The couplings $u$ and $v$ have straightforward identifications in the
$\cN=4$ gauge theory on the $TN$ space. First of all, the parameters $u_i$
are Wilson loops along the circle of the $D4$ compactified on $TN
\times S^1$, and so in the four-dimensional theory they just describe
the values of the scalar fields on the Higgs moduli space. That is,
they parametrize the positions $u_i$ of the $N$ D4-branes along the
$S^1$. Clearly, we are not interested in describing these kind of
configurations where the gauge group $U(N)$ gets broken to $U(1)^N$
(or some intermediate case). Therefore we will in general put
$u=0$. 

The parameters $v_a$ are the Wilson lines on the D6-branes and are
directly related to fluxes along the non-trivial two-cycles of $TN_k$
and (in the limiting case) on the $A_{k-1}$ geometry. To see this, let
us briefly review how the world-volume fields of the D6-branes are
related to the $TN$ geometry in the M-theory compactification. 

First of all, the positions of the NUTs $\vec{x}_a$ of the $TN$ manifold
are given by the vev's of the three scalar Higgs fields of the 6+1
dimensional gauge theory on the D6-brane.  In a similar fashion the
$U(1)$ gauge fields $\wt A_a$ on the D6-branes are obtained from the
3-form $C_3$~field in M-theory. More precisely, if $\w_a$ are the $k$
harmonic two-forms on $TN_k$ introduced in section 2.2, we have a
decomposition
\be
\label{C}
C_3 = \sum_a \w_a \wedge \wt A_a.
\ee
We recall that the forms $\w_a$ are localized around the centers
$\vec{x}_a$ of the $TN$ geometry (the fixed points of the circle
action). So in this fashion the bulk $C_3$~field gets replaced by $k$
$U(1)$~brane fields $\wt A_a$. This relation also holds for a single
D6-brane, because the two-form $\w$ is normalizable in the
$TN_1$ geometry. Relation (\ref{C}) holds in particular for a flat
connection, in which case we get the M-theory background
$$
C_3 = \sum_a v_a\, \w_a \wedge dz + c.c.
$$
Reducing this 3-form down to the type IIA configuration on $TN \times
S^1$ gives a mixture of NS $B$ fields and RR $C_3$ fields on the Taub-NUT 
geometry. Finally, in the $\cN=4$ gauge theory this translates (for an
instanton background) into a topological coupling
$$
\int v \wedge \Tr\, F_+  + \bar{v} \wedge \Tr\, F_-, 
$$
with $v$ the harmonic two-form
$$
v = \sum_a  v_a \w_a.
$$
The existence of this coupling can also be seen by recalling that
the M5-brane action contains the term $\int H \wedge C_3$. On the
manifold $M \times T^2$ the tensor field strength $H$ reduces as $H =
F_+\wedge d\bar{z} + F_- \wedge dz$ and similarly one has $C_3 = v
\wedge dz + \bar{v} \wedge d \bar{z}$, which gives the above
result. If one thinks of the gauge theory in terms of a D3-brane, the
couplings $v, \bar{v}$ are the fluxes of the complexified 2-form
combination $B_{RR} + \tau B_{NS}$.

\subsub{Chiral anomaly}

We should address another point: the chiral fermions on the I-brane
are obviously anomalous. Under a gauge transformation of, say, the
$U(N)$ gauge field
$$
\delta A = D\xi,
$$
the effective action of the fermions transforms as
$$
k \int_{T^2}\Tr(\xi F_A).
$$ 
A similar story holds for the $U(k)$ gauge symmetry. Nonetheless, the
overall theory including both the chiral fermions on the I-brane and
the gauge fields in the bulk of the D-branes is consistent, due to the
coupling between both systems. The consistency is ensured by
Chern-Simons terms in the D-brane actions, which cancel the anomaly
through the process of anomaly inflow \cite{Ibrane,Ibrane5}. For
example, on the D4-brane there is a term coupling to the RR 2-form
(graviphoton) field strength $G_2$:
\be
\label{CS}
I_{CS} = {1 \over 2 \pi} \int_{T^2 \times \R^3} G_2 \wedge CS(A),
\ee
with Chern-Simons term 
$$
CS(A) = \Tr\big(AdA + {2\over 3} A\wedge A \wedge A \big).
$$
Because of the presence of the D6-branes, the 2-form $G_2$ is no
longer closed, but satisfies instead
$$
dG_2 = 2 \pi k \cdot \delta_{T^2}.
$$ 
Therefore under a gauge transformation $\delta A = D\xi$ the D4-brane
action gives the required compensating term
$$
\delta I_{CS} = {1 \over 2 \pi} \int G_2 \wedge d\, \Tr(\xi F_A) =  -k \int_{T^2} \ \Tr(\xi F_A),
$$
which makes the whole system gauge invariant.

\subsection{Conformal embeddings and level-rank duality}

The system of intersecting branes gives an elegant
realization of the level-rank duality
$$
\widehat{su}(N)_k \ \leftrightarrow \ \widehat{su}(k)_N
$$ 
that is well-known in CFT and 3d topological field theory. The
analysis has been conducted in \cite{Ibrane5} for a system of D5-D5
branes, which is of course T-dual to the D4-D6 system that we consider
in this paper. Hence we can follow this analysis to a large extent.

The system of $Nk$ free fermions has central charge $c=Nk$ and gives a
realization of the $\widehat{u}(Nk)_1$ affine symmetry at level
one. In terms of affine Kac-Moody Lie groups we have the embedding
\be
\label{conf}
\widehat{u}(1)_{Nk} \times \widehat{su}(N)_k \times \widehat{su}(k)_N \subset \widehat{u}(Nk)_1.
\ee
This is a conformal embedding, in the sense that the central charges
of the WZW models on both sides are equal. Indeed, using that the
central charge of $\widehat{su}(N)_k$ is
$$
c_{N,k} = \frac{k(N^2-1)}{k+N},
$$
it is easily checked that
$$
1 + c_{N,k} + c_{k,N} = Nk.
$$ 
The generators for these commuting subalgebras are bilinears
constructed out of the fermions $\psi_{i,a}$ and their conjugates
$\psi_{i,a}^\dagger$.  In terms of these fields one can define the
currents of the $\widehat{u}(N)_k$ and $\widehat{u}(k)_N$ subalgebras
as respectively
$$
J_{j \overline k}(z) = \sum_a \psi_{j}{}^a \psi^\dagger_{\overline k a}, 
$$
and
$$
J_{\overline a b}(z) = \sum_j \psi_{j,\overline a} \psi^\dagger_{b}{}^j.
$$

Now it is exactly the conformal embedding (\ref{conf}) that gives the
most elegant explanation of level-rank duality. This correspondence
should be considered as the affine version of the well-known
Schur-Weyl duality for finite-dimensional Lie groups.  Let us recall
that the latter is obtained by considering the (commuting)
actions of the unitary group and symmetric group
$$
U(N) \times S_k \subset U(Nk)
$$ 
on the vector space $\C^{Nk}$, regarded as the $k$-th tensor product
of the fundamental representation $\C^N$.  Schur-Weyl duality is
the statement that the corresponding  group algebras are maximally
commuting in ${\rm End}\left((\C^N)^{\otimes k}\right)$, in the sense
that the two algebras are each other's commutants. 
Under these actions one obtains the decomposition
$$
\C^{Nk} = \bigotimes_\rho V_\rho\otimes \widetilde{V}_\rho,
$$
with $V_\rho$ and $\widetilde{V}_\rho$ irreducible representations of
$u(N)$ and $S_k$ respectively. Here $\rho$ runs over all partitions of
$k$ with at most $N$ parts. This duality gives the famous pairing
between the representation theory of the unitary group and the
symmetric group.

In the affine case we have a similar situation, where we now take the
$k$'th tensor product of the $N$ free fermion Fock spaces, viewed as
the fundamental representation of $\widehat{u}(N)_1$. The symmetric
group $S_k$ gets replaced by $\widehat{u}(k)_N$ (which reminds one of
constructions in D-branes and matrix string theory, where the symmetry
group appears as the Weyl group of a non-Abelian symmetry). The affine
Lie algebras
$$
\widehat{u}(1)_{Nk} \times \widehat{su}(N)_k \times \widehat{su}(k)_N 
$$
again have the property that they form maximally commuting subalgebras
within $\widehat{u}(Nk)_1$. The total Fock space $\cF^{\otimes Nk}$ of $Nk$ free
fermions now decomposes under the embedding (\ref{conf}) as
\be
\label{decom}
\cF^{\otimes Nk} = \bigoplus_{\rho} U_{\|\rho\|} \otimes V_{\rho} \otimes
{\wt V}_{\wt\rho}.
\ee
Here $U_{\|\rho\|}$, $V_{\rho}$ and ${\wt V}_{\wt\rho}$ denote irreducible
integrable representations of $\widehat{u}(1)_{Nk}$, $\widehat{su}(k)_N$, 
and $\widehat{su}(N)_k$ respectively.

The precise formula for the decomposition (\ref{decom}) is a bit
complicated, in particular due to the role of the overall $U(1)$
symmetry, and is given in detail in appendix \ref{app-decompose}.  But
roughly it can be understood as follows: the irreducible
representations of $\widehat{u}(N)_k$ are given by Young diagrams that
fit into a box of size $N \times k$. Similarly, the representations of
$\widehat{u}(k)_N$ fit in a reflected box of size $k \times N$. In
this fashion level-rank duality relates a representation $V_\rho$ of
$\widehat{u}(N)_k$ to the representation ${\wt V}_{\wt\rho}$ of
$\widehat{u}(k)_N$ labeled by the transposed Young diagram. If we
factor out the $\widehat{u}(1)_{Nk}$ action, we get a representation
of charge $\|\rho\|$, which is related to the total number of boxes
$|\rho|$ in $\rho$ (or equivalently $\wt\rho$).

At the level of the partition function we have a similar decomposition
into characters. To write this in more generality it is useful to add
the Cartan generators. That is, we consider the characters for
$\widehat{u}(N)_k$ that are given by
$$ 
\chi_\rho^{\widehat{u}(N)_k} (u,\tau) = \Tr_{\strut V_\rho} \left( e^{2\pi i u_j
  J_0^j} q^{L_0-c_{N,k}/24} \right),
$$
and similarly for $\widehat{u}(k)_N$ we have
$$
\chi_{\wt\rho}^{\widehat{u}(k)_N} (v,\tau) = \Tr_{\strut \wt V_{\wt\rho}} 
\left( e^{2\pi i v_a J_0^a} q^{L_0-c_{k,N}/24} \right).
$$ 
Here the diagonal currents
$$
J_0^j = \oint \frac{dz}{2\pi i} J_{jj}(z),\qquad  
J_0^a = \oint \frac{dz}{2\pi i} J_{aa}(z)
$$ 
generate the Cartan tori $U(1)^N \subset U(N)$ and $U(1)^k \subset U(k)$.

Including the Wilson lines $u$ and $v$ for the $U(N)$ and $U(k)$ gauge
fields, the partition function of the I-brane system is given by the
character of the fermion Fock space 
\bea
Z_I(u,v,\tau) & = & 
\Tr_{\strut \mathcal{F}}\Big(e^{2\pi i (u_jJ^j_0 + v_a J^a_0)}\, q^{L_0 - \frac{Nk}{24}}
\Big) \\
& = & 
q^{-\frac{Nk}{24}}\prod_{\scriptstyle j=1, \ldots,N \atop 
\scriptstyle a=1,\ldots,k} \prod_{n \geq 0}\Big(1+e^{2\pi i (u_j+v_a)}
q^{n+1/2} \Big)  
\Big(1+e^{- 2\pi i (u_j+v_a)} q^{n+1/2} \Big). \nonumber
\eea
Writing the decomposition (\ref{decom}) in terms of characters
gives 
$$
Z_I(u,v,\tau) = \sum_{[\rho]\subset
  \mathcal{Y}_{N-1,k}} \sum_{j=0}^{N-1} \sum_{a=0}^{k-1}
\chi^{\widehat{u}(1)_{Nk}}_{|\rho|+jk+aN}(N |u| + k
|v|,\tau)\ \chi^{\widehat{su}(N)_k}_{\sigma^j_N(\rho)} (\overline{u},\tau)
\chi^{\widehat{su}(k)_N}_{\sigma^a_k(\wt \rho)}
(\overline{v},\tau),
$$
where the Young diagrams $\rho\in \cY_{N-1,k}$ of size $(N-1)\times k$
represent $\widehat{su}(N)_k$ integrable representations and $\sigma$
denote generators of the outer automorphism groups $\Z_N$ and $\Z_k$
that connect the centers of $SU(N)$ and $SU(k)$ to the $U(1)$
factor (see again the appendix for notation and more details).

\subsection{Deriving the McKay-Nakajima correspondence}   \label{ssec-mckay}

In the intersecting D-brane configuration both the D4-branes and the
D6-branes are non-compact. So, we can choose both the $U(N)$ and
$U(k)$ gauge groups to be non-dynamical and freeze the background
gauge fields $A$ and $\wt A$. In fact, this set-up is entirely
symmetric between the two gauge systems, which makes level-rank
duality transparent. 

However, in order to make contact with the $\cN=4$ gauge theory
computation, we will have to break this symmetry. Clearly, we want the
$U(N)$ gauge field to be dynamical --- our starting point was to
compute the partition function of the $U(N)$ Yang-Mills theory. The
$U(k)$ symmetry should however {\it not} be dynamical, since we want
to freeze the geometry of the Taub-NUT manifold. So, to derive the
gauge theory result, we will have to integrate out the $U(N)$ gauge
field $A$ on the I-brane. Particular attention has to be payed to the
$U(1)$ factor in the CFT on the I-brane. We will argue that in this
string theory set-up we should not take that to be dynamical.

Therefore we are dealing with a partially gauged CFT or coset theory
$$
\widehat{u}(Nk)_1/\widehat{su}(N)_k.
$$
In particular the $\widehat{su}(N)_k$ WZW model will be replaced by the
corresponding $G/G$ model. Gauging the model will reduce the
characters. (Note that this only makes sense if the Coulomb parameters
$u$ are set to zero. If not, we can only gauge the residual gauge
symmetry, which leads to fractionalization and a product structure.)
In the gauged WZW model, which is a topological field theory, only the
ground state remains in each irreducible integrable representation. So
we have a reduction
$$
\chi^{\widehat{su}(N)_k}_\rho(\overline{u},\tau)\ \to\ q^{h_{\rho}-c/24},
$$ 
with $h_\rho$ the conformal dimension of the ground state
representation $\rho$. Note that the choice of $\rho$ corresponds
exactly to the boundary condition for the gauge theory on the
$A_{k-1}$ manifold. We will explain this fact, that is crucial to the
McKay correspondence, in a moment.

Gauging the full I-brane theory and restricting to
the sector $\rho$ finally gives
$$
Z_I(u,v,\tau)\ \to\ Z^{N,k}_\rho(v,\tau) = q^{h_{\rho}-c/24}
\sum_{a=0}^{k-1}
\chi^{\widehat{u}(1)_{Nk}}_{|\rho|+aN}(k|v|,\tau)\ 
\chi^{\widehat{su}(k)_N}_{\sigma^a_k(\wt\rho)}(\overline{v},\tau).
$$ 
Up to the $\chi^{\widehat{u}(1)_{Nk}}$ factor, this reproduces the
results presented in \cite{vafa-witten,nakajima} for ALE spaces, which
involve just $\widehat{su}(k)_N$ characters. This extra factor is
is due to additional monopoles mentioned in section
\ref{ssec-realization}. They are related to the finite radius $S^1$
at infinity of the Taub-NUT space and are absent in case of ALE
geometries. 

In fact, the extra $U(1)$ factor can already be seen at the classical level,
because the extra normalizable harmonic two-form $\omega$ in
(\ref{omega}) disappears in the decompactification limit where
$TN_k$ degenerates into $A_{k-1}$. The lattice $H^2(TN_k,\Z)$ is isomorphic 
to $\Z^k$ with the standard inner product and contains the root
lattice $A_{k-1}$ as a sublattice given by $\sum_I n_I=0$. Note also
that the lattice $\Z^k$ is not even, which explains why the I-brane
partition function has a fermionic character and only transforms under
a subgroup of $SL(2,\Z)$ that leaves invariant the spin structure on
$T^2$.

\subsub{Relating the boundary conditions}

By relating the original four-dimensional gauge theory to the
intersecting brane picture one can in fact derive the McKay
correspondence directly. Moreover we can understand the appearance of
characters of the WZW models (for both the $SU(N)$ and the $SU(k)$
symmetry) in a more natural way in this set-up.  Recall that the
$SU(N)$ gauge theory on the $A_{k-1}$ singularity or $TN_k$ manifold is
specified by a boundary state. This state is given by picking a flat
connection on the boundary that is topologically $S^3/\Z_k$.  If we
think of this system in radial quantization near the boundary, where
we consider a wave function for the time evolution along
$$
S^3/\Z_k \times \R,
$$ we have a Hilbert space with one state $|\rho\rangle$ for each
$N$-dimensional representation
$$
\rho:\ \Z_k \to U(N).
$$ 

After the duality to the I-brane system, we are dealing with a
five-dimensional $SU(N)$ gauge theory on $\R^3 \times T^2$, with 
$k$ D6-branes intersecting it along $\{p\}\times T^2$ where $p$ is
(say) the origin of $\R^3$. Here the boundary of the D4-brane system
is $S^2 \times T^2$.  In other words, near the boundary the space-time
geometry looks like $\R\times S^2\times T^2$. We now ask ourselves what
specifies the boundary states for this theory.  Since we need a finite
energy condition, this is equivalent to considering the IR limit of the
theory.
In M-theory the $S^1$-bundle over 
$S^2$ carries a first Chern class $k$, which translates into the flux
of the graviphoton field strength 
$$
\int_{S^2} G_2 = 2\pi k.
$$
Therefore the term 
$$
\int_{S^2\times T^2\times \R} G_2\wedge CS(A),
$$
living on the D4 brane, leads upon reduction on $S^2$ (as is done in
\cite{Ibrane5}) to the term
$$
I_{CS} = 2\pi k \int_{T^2 \times \R}CS(A).
$$ 
Hence we have learned that the boundary condition for the D4-brane
requires specifying a state of the $SU(N)$ Chern-Simons theory at
level $k$ living on $T^2$.  The Hilbert space for Chern-Simons theory
on $T^2$ is well-known to have a state for each integrable
representation of the $\widehat{u}(N)_k$ WZW model, which up to the
level-rank duality described in the previous section, gives the McKay
correspondence.  

In fact, the full level-rank duality can be brought to
life. Just as we discussed for the $N$ D4-branes, a $SU(k)$ gauge theory 
lives on the $k$ D6-branes on $T^2\times \R^5$.  The boundary of
the space is $S^4\times T^2$. Furthermore, taking into account that the $N$
D4-branes source the $G_4$ RR flux through $S^4$, we get, as in the
above, a $SU(k)$ Chern-Simons theory at level $N$ living on
$T^2\times \R$.  Therefore the boundary condition should be specified by a
state in the Hilbert space of the $SU(k)$ Chern-Simons theory on
$T^2$.  So we see three distinct ways to specify the boundary
conditions: as a representation of $\Z_k$ in $SU(N)$, as a character
of $SU(N)$ at level $k$, and as a character of $SU(k)$ at level $N$.
Thus we have learned that, quite independently of the fermionic realization,
there should be an equivalence between these objects.

To make the map more clearly we could try to show that the choice of the
flat connection of the $SU(N)$ theory on $S^3/ \Z_k$ gets mapped to
the characters that we have discussed in the dual intersecting brane
picture.  To accomplish this, recall that the original $SU(N)$ action
on the $A_{k-1}$ space leads to a boundary term (modulo an integer
multiple of $2\pi i \tau$) given by the Chern-Simons invariant
$$
 {\tau\over 4\pi i}\int_{A_{k-1}} \Tr\,F \wedge F =
{\tau\over 4\pi i} \int_{S^3/ \Z_k} CS(A).
$$ 
Restricting to a particular flat connection on $S^3/ \Z_k$ yields
the value of the classical Chern-Simons action.

If we show that
$$
S(\rho) = {1\over 8\pi^2} \int_{S^3/ \Z_k} CS(A)
$$ 
for the flat connection $\rho$ on $S^3/\Z_k$ gets mapped to the
conformal dimension $h_\rho$ of the corresponding state of the quantum
Chern-Simons theory on $T^2$, we would have completed a direct check of the map, 
because the gauge coupling constant $\tau$ above is nothing but the 
modulus of the torus in the dual description.

To see how this works, let us first consider the abelian case of
$N=1$. In that case the flat connection $\rho$ is given by a phase
$e^{2\pi i n/k}$ with $n \in \Z/k\Z$. The corresponding CS term gives
$$
S^{U(1)}(\rho)= {n^2 \over 2k}.
$$
This is the conformal dimension of a primary state of the
$U(1)$ WZW model at level $k$. 

A general $U(N)$ connection can always be diagonalized to $U(1)^N$,
which therefore gives integers $n_1,\ldots,n_N \in \Z/k\Z$. The
Chern-Simons action is therefore given by
$$
S^{U(N)}(\rho) = \sum_{I=1}^N {n_I^2 \over 2k}.
$$ 
On the other hand, a conformal dimension of a primary state in the
corresponding WZW model is given by
$$
h_{\rho} = \frac{C_2(\rho)}{2(k+N)},
$$
where $\rho$ is an irreducible integrable $\widehat{u}(N)_k$
weight. Such a weight can be encoded in a Young diagram with at most
$N$ rows of lengths $R_I$. There is a natural change of basis $n_I =
R_I + \rho^{Weyl}_I$ where we shift by the Weyl vector
$\rho^{Weyl}$. If we decompose $U(N)$ into $SU(N)$ and $U(1)$, the
basis $n_I$ cannot be longer than $k$, which relates to the condition $n_I
\in \Z_k$ on the Chern-Simons side. In this basis the second Casimir $C_2$ takes a
simple form. Therefore the conformal dimension becomes
$$
h_{\rho} = -\frac{N(N^2-1)}{24(k+N)} + \frac{1}{2(k+N)} \sum_{I=1}^N n_I^2.
$$
The constant term combines nicely with the central charge contribution
$-c_{N,k}/24$ to give an overall constant $(N^2-1)/24$. Apart from
this term we see that $h_{\rho}$ indeed matches the expression for
$S^{U(N)}(\rho)$ given above, up to the usual quantum shift $k \to
k+N$.

According to the McKay correspondence one might expect to find a
relation between representations of $\Z_k$ and $\widehat{u}(k)_N$
integrable weights. Instead, we have just shown how $\widehat{u}(N)_k$
weights $\rho$ arise. Nonetheless, one can relate integrable weights
of those algebras by a transposition of the corresponding Young
diagrams. Then the conformal dimensions of $\widehat{u}(k)_N$ weights
$\wt\rho$ are determined by the relation \cite{nrs}
$$
h_{\rho}+h_{\wt\rho} = \frac{|\rho|}{2} - \frac{|\rho|^2}{2Nk},
$$
which is a consequence of the level-rank duality described in appendix
\ref{app-decompose}.  The above chain of arguments connects $\Z_k$
representations and $\widehat{u}(k)_N$ integrable weights, thereby
realizing the McKay correspondence.

\subsection{Orientifolds and $SO/Sp$ gauge groups}

So far we have considered a system of $N$ D4-branes and $k$
D6-branes intersecting along a torus, whose low energy theory is
described by $U(N)$ and $U(k)$ gauge theories on each stack of branes,
together with bifundamental fermions. We can reduce this system to
orthogonal or symplectic gauge groups in a standard way by adding an
orientifold plane. This construction can also be lifted to
M-theory. Let us recall that D6-branes in our system originated from
a Taub-NUT solution in M-theory. The O6-orientifold can also be
understood from M-theory perspective, and it corresponds to the
Atiyah-Hitchin space \cite{sen-M}. Combining both ingredients, it is
possible to construct the M-theory background for a collection of
D6-branes with an O6-plane. The details of this construction are
explained in \cite{sen-M}.

Let us see what are the consequences of introducing the orientifold
into our I-brane system. We start with a stack of $k$ D6-branes. To
get orthogonal or symplectic gauge groups one should add an
orientifold O6-plane parallel to D6-branes \cite{gim-pol}, which
induces an orientifold projection $\Omega$ which acts on the
Chan-Paton factors via a matrix $\gamma_{\Omega}$. Let us recall there
are in fact two species O6${}^{\pm}$ of such an orientifold. As the
$\Omega$ must square to identity, this requires 
$$
\gamma_{\Omega}^t = \pm \gamma_{\Omega},
$$ 
with the $\pm$ sign corresponding to O6$^{\pm}$-plane, which gives
respectively $SO(k)$ and $Sp(2k)$ gauge group. In the former case $k$
can be even or odd; $k$ odd requires having \emph{half-branes}, fixed
to the orientifold plane (as explained {\it e.g.}~in
\cite{ev-clif-shap}). 

Let us add now $N$ D4-branes intersecting D6 along two directions. The
presence of O6${}^{\pm}$-plane induces appropriate reduction of the D4
gauge group as well. The easiest way to argue what gauge group arises
is as follows. We can perform a T-duality along three directions to get a
system of D1-D9-branes, now with a spacetime-filling O9-plane. This is
analogous to the D5-D9-09 system in \cite{gim-pol}, in which case
the gauge groups on both stacks of branes must be different (either
orthogonal on D5-branes and symplectic on D9-branes, or the other
way round); the derivation of this fact is a consequence of having 4
possible mixed Neumann-Dirichlet boundary conditions for open strings
stretched between branes. On the contrary, for D1-D9-O9 system
there is twice as many possible mixed boundary conditions, which in
consequence leads to the same gauge group on both stack of branes. By
T-duality we also expect to get the same gauge groups in D4-D6 system
under orientifold projection. 

Let us explain now that the appearance of the same type of gauge
groups is consistent with character decompositions resulting from
consistent conformal embeddings or the existence of the so-called dual
pairs of affine Lie algebras related to systems of free fermions.  We
have already come across one such consistent embedding in (\ref{conf})
for $\widehat{u}(Nk)_1$. A dual pair of affine algebras in this case
is $(\widehat{su}(N)_k,\, \widehat{su}(k)_N)$. These two algebras are
related by the level-rank duality discussed in the appendix
\ref{app-decompose}.  As proved in \cite{jimbo-miwa,Hasegawa}, all
other consistent dual pairs are necessarily of one of the following
forms
$$ (\widehat{sp}(2N)_k ,\, \widehat{sp}(2k)_N), $$
$$ (\widehat{so}(2N+1)_{2k+1} ,\, \widehat{so}(2k+1)_{2N+1}), $$
$$ (\widehat{so}(2N)_{2k+1} ,\, \widehat{so}(2k+1)_{2N}),  $$
$$ (\widehat{so}(2N)_{2k} ,\, \widehat{so}(2k)_{2N}). $$ 
Corresponding expressions in terms of characters, analogous to
(\ref{uNk-decompose}), are also given in \cite{Hasegawa}. The crucial
point is that both elements of those pair involve algebras of the same
type, which confirms and agrees with the string theoretic orientifold
analysis above.

Finally we wish to stress that the appearance of $U$, $Sp$ and $SO$
gauge groups which we considered so far in this paper is related to
the fact that their respective affine Lie algebras can be realized in
terms of free fermions, which arise on the I-brane from our
perspective.  It turns out there are other Lie groups $G$ whose affine
algebras have free fermion realization.  There is a finite number of
them, and fermionic realizations can be found only if there exists a
symmetric space of the form $G'/G$ for some other group $G'$
\cite{gno}.  It is an interesting question whether I-brane
configurations could be engineered in string theory which would
support fermions realizing all those affine algebras, and what would
be a physical interpretation of the corresponding symmetric
spaces.

From a geometric point of view we can remark the following. For ALE
singularities of $A$-type and $D$-type a non-compact dimension can be
compactified on a $S^1$ to give Taub-NUT geometries. For exceptional
groups such manifolds do not exist. But one can compactify \emph{two}
directions on a $T^2$ to give an elliptic fibration. In this setting
exotic singularities can appear as well. Such construction have a direct
analogue in type IIB string theory where they correspond to a
collection of $(p,q)$ 7-branes \cite{vafa-morrison}. The I-brane is
now generalized to the intersection of $N$ D3-branes with this
non-abelian 7-brane configuration \cite{D3D7O}. However, there is in general
no regime where all the 7-branes are weakly coupled, so it is not
straightforward to write down the I-brane system.

\section{$\cN=2$ gauge theories and curved I-branes}
  
\label{sec-strings}

There is no reason why the above argument relating gauge theories on
an $A_{k-1}$ singularity to an I-brane system produced by $k$
intersecting D6-branes, cannot be restricted to the trivial case of an
$A_0$ ``singularity''. That is, with the same methods we can try to
compute the partition function of the supersymmetric gauge theory on
$\R^4$ by embedding it into a $TN_1$ manifold, which we can then think
of as interpolating smoothly between $\R^4$ and $\R^3 \times S^1$,
depending on the value of its compactification radius $R$. Of course,
in this duality we do take the size of a circle in M-theory from large
to small, and therefore the relation between the gauge theory and the
I-brane system should be considered as a strong-weak duality in string
theory.

For example, we could consider the simplest possible case of a $U(1)$
gauge theory corresponding to a single D4-brane on $\R^4$. Using the
duality with I-branes this theory would be then mapped to a single
chiral fermion $\psi(z)$ on the I-brane. Note that, for the
appropriate values of the background moduli, this $U(1)$ gauge theory
has a stringy nature and there are still point-like instantons in this
model (bound states with D0-branes), which explains why this partition
function is non-trivial function of the gauge coupling $\tau$:
$$ 
Z(\tau) = {\theta_3(v,\tau) \over \eta(\tau)} = \sum_{n\in\Z} {e^{\pi
    i \tau n^2 + 2\pi i v n} \over \eta(\tau)},
$$
with
$$ 
\eta(\tau) = q^{1/24} \prod_{n>0} \left(1-q^n \right).
$$ 

Up to now we have only been considering I-branes with worldsheet
$T^2$. However, we can generalize this easily to more general
topologies. In this fashion we will naturally relate to gauge theories
with less than maximal supersymmetry.

\subsection{I-branes on general curves}

We will now turn to gauge theories with $\cN=2$ and $\cN=1$
supersymmetry. It is well-known that these models can be engineered in
string theory by considering more complicated compactifications and
brane configurations. In fact, in many ways the most elegant starting
point is an M5-brane configuration in M-theory.

Let us start with the compactification
$$
\hbox{(M)}\quad TN \times \B \times \wt\R^3,
$$ 
corresponding to the top box in fig. 1 (with $S^1$ decompactified).
Here we have denoted the three non-compact directions as $\wt\R^3$
to distinguish them from the $\R^3$ in the base of $TN$. We further
pick $\B$ to be a flat complex surface that is topologically a $T^4$ or
some decompactification of it.  
That is, in the most general case $\B$ will be a product
$$
\B = E \times E'
$$ 
of two elliptic curves. But more often we will consider the
degenerations $\B = \C^* \times \C^*$ and $\B = \C \times \C$, or any
mixed combination. (In the relation with integrable hierarchies the
cases $\C,$ $\C^*$, and $E$ correspond to rational, trigonometric, and
elliptic solutions respectively.) We will denote the affine
coordinates on $\B$ as $(x,y) \in \B$. The complex surface $\B$ has a
(2,0) holomorphic form
$$
\omega = dx \wedge dy.
$$
We will now pick a holomorphic curve $\Sigma$ inside $\B$ given by an equation
$$
\Sigma:\ P(x,y)=0,
$$
and wrap for the moment a single M5-brane over $TN \times \Sigma$.
Because $\Sigma$ is holomorphically embedded this is a
configuration with $\cN=2$ supersymmetry in four dimensions.

Now there are two obvious reductions to type IIA string theory
depending on whether we take the $S^1$ in the Taub-NUT fibration,
or an $S^1$ inside $\B$. In the first case we obtain the Type IIA
compactification of the form
$$
\hbox{(IIA)}\quad \R^3 \times \B \times\wt \R^3.
$$
Again there will be a D4-brane wrapping $\R^3 \times \Sigma$ with
$\Sigma \subset \B$, together with $k$ D6-branes wrapping $\B \times
\wt\R^3$. In this case the I-brane is completely wrapping the curve
$\Sigma$. This configuration is illustrated in fig. 3.

\begin{figure}[t]
\begin{center}
\includegraphics[width=6cm]{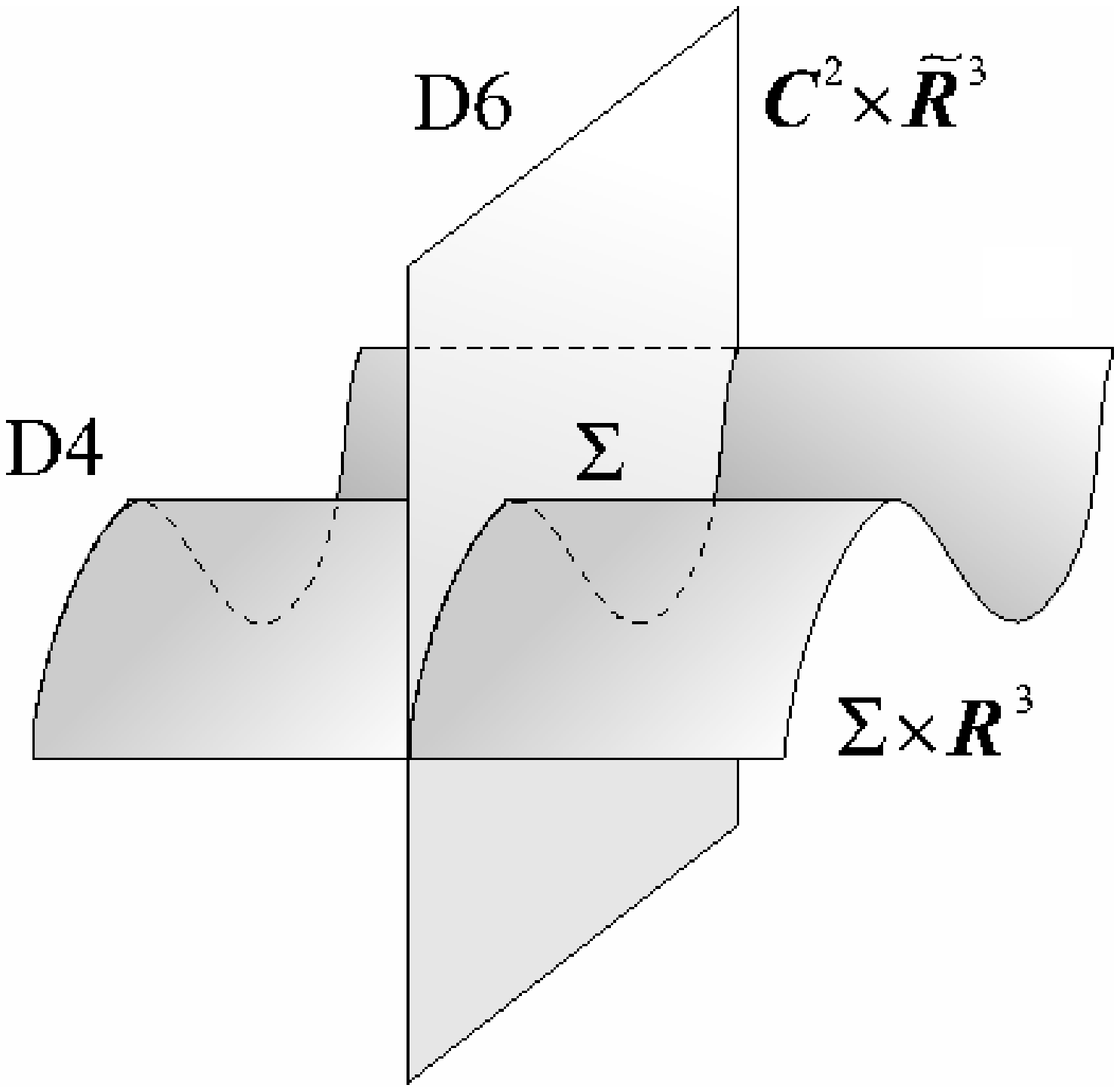}\\[5mm]
\parbox{12cm}{\small \bf Fig.\ 3: \it A more general configuration of
  D4 and D6-branes where the intersection locus is an affine holomorphic curve $\Sigma$.}
\end{center} 
\end{figure}

In this generalized geometry we should consider the free fermion
system on a higher genus Riemann surface with action
$$
I = \int_\Sigma \psi^\dagger \overline\partial \psi.
$$

In the dual interpretation we will compactify $\B$ along a $S^1$ down
to a three dimensional base $\cB_3$. The curve $\Sigma$, and therefore
also the M5-brane, will partially wrap this $S^1$. Consequently, we
arrive at a configuration of NS 5-branes and D4-branes that are
spanned between them \cite{M-4d}. In the classical situation discussed by Witten
we take $\B = \C \times \C^*$ and end up with a IIA string theory on
$$ 
\hbox{(IIA)}\quad TN \times \R^6
$$ 
with a set of parallel NS 5-branes with D4-branes ending on them. This
is exactly the brane configuration that engineers $\cN=2$ gauge
theories.

In this case we can again compare the I-brane with the gauge theory
computation. In the gauge theory we are computing two
contributions. Firstly, there is a
gauge coupling matrix $\tau_{IJ}$ of the $U(1)^g$ fields $F^I$
$$
\int_{TN} {i \over 4 \pi} \tau_{IJ} F^I_+ \wedge F^J_+ +
v_I \wedge F^I_+.
$$ 
for a genus $g$ curve $\Sigma$. On the $TN$ geometry the gauge field strengths $F^I$ have fluxes in
the lattice 
$$
[F^I/2\pi] = p^I \in H^2(TN,\Z).
$$ 
Since the cohomology lattice $H^2(TN,\Z) \cong \Z^k$, these fluxes are
labeled by integers $p^I_a$ with $I=1,\ldots, g$ and $a=1,\ldots,k$.

Secondly, there is a gravitational coupling $\cF_1$ that appears in
the term
$$
\int_{TN} \cF_1(\tau)\ \Tr\, R_+ \wedge R_+.
$$
Since the regularized Euler number of $TN_k$ equals $k$, 
combining these two terms yields the partition function
\be
\label{Z-bos}
Z_{gauge} = \sum_{p^I_a \in \Z} e^{\pi i p^I_a \tau_{IJ} p^{J,a} +
  2\pi i v_I^a p^I_a} e^{k \cF_1}.
\ee

In the I-brane model the partition function $Z_{gauge}$ is nothing but
the determinant of the chiral Dirac operator acting on $k$ free
fermions living on the ``spectral curve'' $\Sigma_g$, coupled to the
flat $U(k)$ connection $v$ on corresponding rank $k$ vector bundle
${\cal E} \to \Sigma$. So we are led in a very direct way to
\be
\label{Z-fer}
Z_{gauge}
 = \det \overline\partial_{\cal E}.
\ee
The results (\ref{Z-bos}) and (\ref{Z-fer}) are just the usual
bosonization formula, where the fermion determinant is equivalent to
 a sum over the lattice of momenta together with a boson
determinant. Here we use the identification
$$
\cF_1 = - {1\over 2} \log \det \Delta_\Sigma.
$$

To complete this map we need to show why the $p^I$ are identified with
fermion currents on the Riemann surface through the corresponding
cycle, but this is relatively clear. Consider a cycle $\alpha_I$ on
the Riemann surface and a disc ending on this cycle (which can
always be done as $\Sigma$ is contractible in the full CY).  Then the
statement that $F^I$ is turned on corresponds to the fact that the
integral of the corresponding flux over this disc is not zero.  Since
the fermions are charged under the $U(k)$ gauge group, this means that
they pick up a phase as they go along this cycle on the Riemann
surface (the Aharanov-Bohm effect).  Thus the holonomy of the fermions
correlates with the $p^I$.  Later (in section 3.4), we provide an alternative  
view of the fluxes $p^I$:
they also correspond to D4-branes, wrapping 4-cycles of the Calabi-Yau and
bound to the D6-brane.

\subsection{The duality chain}  \label{ssec-chain}

Holomorphic quantities in supersymmetric gauge theories are
intimately connected to topological string amplitudes on the
Calabi-Yau geometry that engineer the gauge system. We now want to
connect our results to topological strings by going through another
chain of dualities, represented by the vertical sequence of boxes in fig. 1. 
For this we consider a slightly more general
compactification of M-theory, namely
$$ 
\hbox{(M)}\quad TN \times \B \times \R^2 \times S^1,
$$ 
corresponding to the top box in fig.~1.
The extra $S^1$ does not really influence the earlier results, since
the D6-branes remain non-compact and therefore the gauge theory that
they support stays non-dynamical. Hence the I-brane configuration remains
the same. In the other compactification with an ensemble of NS
5-branes and D4-branes we can now perform a T-duality on $S^1$ to give
a web of $(p,q)$ 5-branes in Type IIB, which is another familiar and
convenient realization of the $\cN=2$ system.

However, in this situation there is an obvious third possible compactification to
type IIA, by just reducing on the extra $S^1$ that we have
introduced. This will give IIA on
$$
\hbox{(IIA)}\quad TN \times \B \times \R^2
$$ 
with $N$ NS 5-branes wrapping $TN \times \Sigma$. We haven't gained
much in this step, but now we can T-dualize the NS 5-brane away to
remain with a purely geometric situation \cite{ns5-ale,self-dual-string}. 
In general a T-duality {\it transverse} to a set of $N$ NS 5-branes
produces a local $A_{N-1}$ singularity of the form
\be
uv = z^N.
\label{tdual}
\ee 
In the case of a single 5-brane $N=1$ this gives an $A_0$
``singularity''.  We recall that world-sheet instanton effects are
important to understand this very non-trivial duality
\cite{ns5-ale,NS5-Tong,NS5-HJ}.

Applying this T-duality in the present set-up gives us a type IIB
compactification of the form
$$
\hbox{(IIB)}\quad TN \times X,
$$
where $X$ is now a non-compact Calabi-Yau geometry of the form
$$
X:\ uv + P(x,y)=0.
$$ 
This can be regarded as a $\C^*$ fibration over $\cB$ where the fibers
degenerate to a pair of intersecting lines $uv=0$ over the locus
$\Sigma$ given by $P(x,y)=0$. This is just an application of
(\ref{tdual}) in the case $N=1$, where $z$ is the local coordinate
transverse to the curve $\Sigma$.

Once we are in this completely geometric phase, without any further
branes, further dualities can bring us into other familiar, but fully
equivalent configurations. Mirror symmetry gives us the background
$$
\hbox{(IIA)}\quad TN \times \wt X,
$$
where the mirror CY geometry $\wt X$ has a toric description. 

Finally we can go up again to eleven-dimensional M-theory
$$
\hbox{(M)}\quad TN \times \wt X \times S^1.
$$ 
This is actually the situation considered in \cite{gv,gsy,dvv} where
five-dimensional black hole degeneracies were computed. 

In fact, we can close the chain of dualities, by reducing once
more to type IIA on the $S^1$ fiber of the $TN$ to obtain
$$
\hbox{(IIA)}\quad \R^3 \times \wt X \times S^1,
$$
where now we have $k$ D6-branes wrapping $\wt X \times S^1$. This is
the situation where one computes Donaldson-Thomas invariants, which
can be viewed as BPS bound states of D0-D2-branes to the D6-brane
\cite{foam,mnop}, at least when the background moduli are in the right
regime for these bound states to exist \cite{denef-moore}.

\subsub{Fermion charges and $U(1)$ fluxes}

It might be good to follow the fermion numbers $p^1,\ldots,p^g$
(through the handles of the Riemann surface) and the dual $U(1)$
holonomies $v_1,\ldots,v_g$ (that couple to these fluxes) through this chain
of dualities. We pick $k=1$ for simplicity. First we remark that we
have to choose a basis of $A$-cycles on $\Sigma$ to define these
quantities.

In the IIB compactification on $TN \times X$ the quantum numbers $p^I$
appear as fluxes of the RR field $G_5$ through a basis of 3-cycles of
the Calabi-Yau $X$. In reduction of $G_5$ to the four-dimensional low-energy
theory on $TN$ this gives the $U(1)$ gauge fields $F^I$ of the vector
multiplets that appear in the gauge coupling
$$
\int_{TN} \tau_{IJ} F^I_+ \wedge F^J_+.
$$

After mapping this configuration to IIA theory on $\R^3 \times S^1
\times \wt X$ with a D6-brane wrapping $S^1 \times \wt X$, the flux
$p$ is carried by the $U(1)$ gauge field strength $F$ on the
world-volume of the D6-brane, $[F^I / 2\pi] = p^I \in H^2(\wt X,\Z)$.
This can be interpreted as a bound state of a single D6-brane with $p$
D4-branes. In a similar fashion the Wilson lines $v_I$ are mapped to
potentials for the D4-branes. So in this duality frame we can simply
shift the fermion number by adding D4-branes.
  
\subsection{Fermions and BPS states}

Perhaps it is clarifying to compare the chiral fermions, that appear
so naturally in the intersecting brane system, to the usual BPS
states. This is most naturally done in the M-theory picture, where we
consider a M5-brane with topology $TN_k \times \Sigma$. 

\begin{figure}[t]
\begin{center}
\includegraphics[width=8cm]{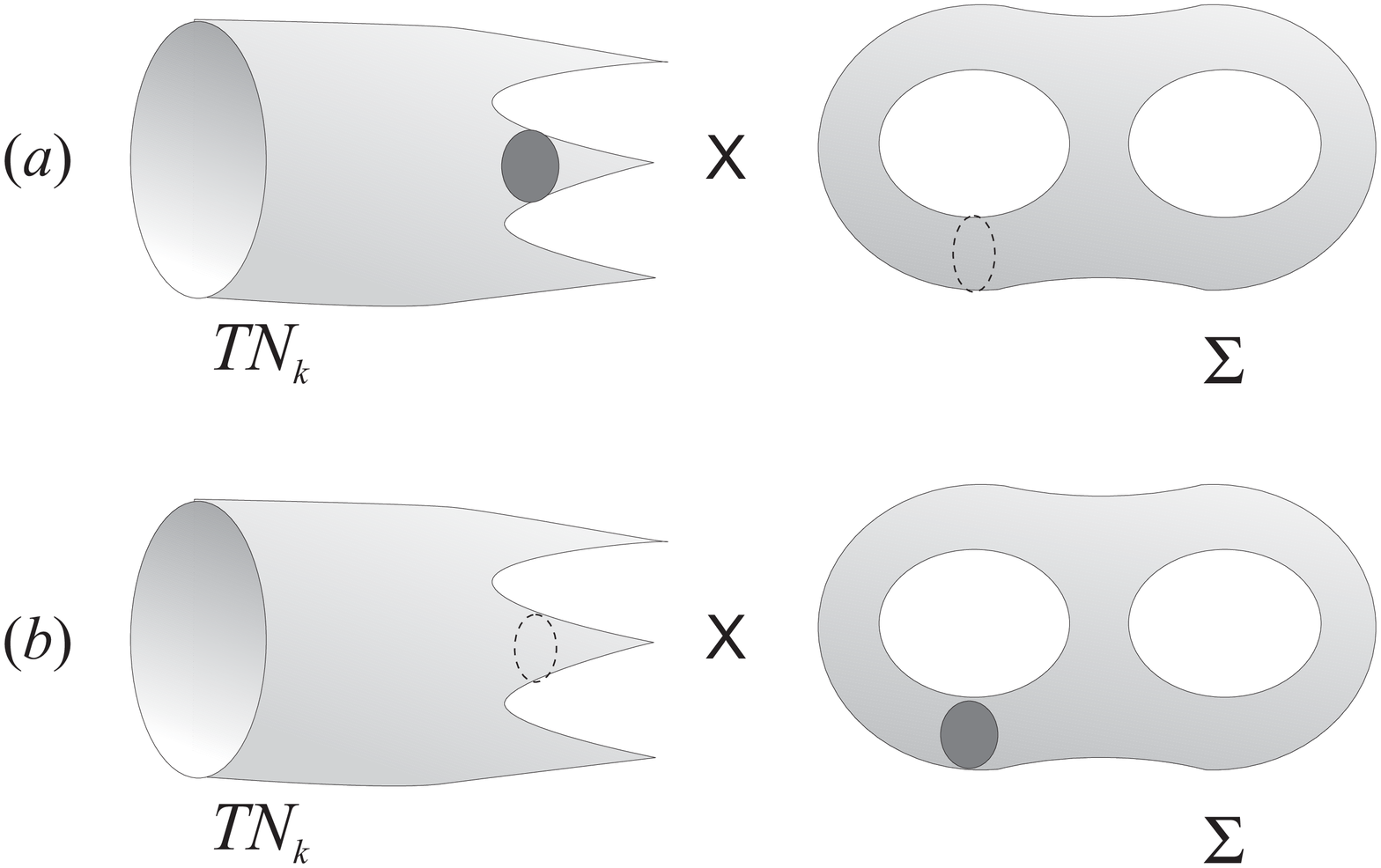}\\[5mm]
\parbox{12cm}{\small \bf Fig.\ 4: \it Two kinds of open M2-brane
  instantons that contribute to a M5-brane with geometry $TN_k \times
  \Sigma$: $(a)$ the free fermions, massless at the NUTs of $TN_k$;
  $(b)$ the usual BPS states that become massless when the Riemann
  surface $\Sigma$ pinches.}
\end{center} 
\end{figure}

First of all, the chiral fermions are given by the open fundamental
strings that stretch between D4-branes and D6-branes. These strings are
lifted to open M2-branes in M-theory. The topology of these
membranes is a two-dimensional disc $D^2$, whose boundary $S^1$ lies
on the M5-brane. It will encircle the $S^1$ of the Taub-NUT
geometry. One way to see this is that the BPS mass of these open M2-brane
states is given by 
$$
Z = \int_{D^2} \w = \oint_{S^1} \eta,
$$
where $\eta$ is the one-form (\ref{eta}) on $TN$. This mass
goes to zero exactly when the M2-brane approaches one of the
NUTs of the Taub-NUT geometry. There the chiral fermions appear. The
time trajectory of these objects will lie on the Riemann surface
$\Sigma$. So, if we compute a fermion one-loop diagram, this is
represented in M-theory in terms of open M2-brane world-volumes with
geometry $D^2 \times S^1$ and boundary $T^2=S^1 \times S^1$ on the
M5-brane. Note that the $S^1$ on the $TN$ factor is filled in by a
disc. This is illustrated in fig. 4a.

On the other hand we have the traditional BPS states in $\cN=2$
compactifications. For example, in the IIB compactification on $TN_k
\times X$ these will be given by D3-branes that wrap special
Lagrangian 3-cycles in $X$. After the duality that maps this
compactification to a M5-brane, these states become open
M2-branes with the topology of a disc as well. But now the boundary $S^1$ of
these discs will lie on the surface $\Sigma$. The time trajectory will
be along the space-time $TN_k$. More invariantly, we have again
M2-branes with world-volumes $D^2 \times S^1$ and boundary $S^1 \times
S^1$, but now the $S^1$ on the Riemann surface $\Sigma$ is filled
in. This makes sense, since the mass of the BPS states, given by
$$
Z= \oint y dx,
$$
goes to zero exactly when the surface $\Sigma$ is pinched or forms a
long neck. These states are the M-theory interpretation of the
well-known massless monopoles of Seiberg and Witten
\cite{M-4d,M-BPS-1,M-BPS-2}.

In a full quantum theory of the M5-brane, both kinds of open M2-branes
should contribute to the partition function. In fact, the boundaries
of these M2-branes are the celebrated self-dual strings that should
describe the M5-brane world-volume theory
\cite{duff-lu,self-dual-string}.  Clearly the corresponding massless
states are contributing in different regimes. In that sense the
relation between the free fermions and the usual BPS states can be
considered as a strong-weak coupling duality.

Chiral fermions localized on the curve $\Sigma$ have also appeared in
the B-model topological string theory on $\R^4 \times X$ \cite{adkmv},
in the context of topological vertex.  It is natural to ask if these
are the same fermions as we have encountered in this paper. An
important property of these fermions in the context of topological
vertex is that the insertions of the operators $\psi(x_i)$, that
create fermions (which is of course not the same as the quanta of the
corresponding field), change the geometry of the curve. The fermions
will produce extra poles in the meromorphic differential $ydx$ which
encodes the embedding of the curve as $P(x,y)=0$.  We have the
identification $y=\partial\phi(x)$, and by bosonization the operator
product with a fermion insertion gives a single pole at each fermion
insertion
$$
\partial\phi(x) \cdot \psi(x_i) \sim {1\over x-x_i} \psi(x_i).
$$

In the superstring such a correlator 
$$
\langle \psi(x_1) \cdots \psi(x_n)\rangle
$$
of fermion creation operators corresponds to the insertion of
D5-branes in type IIB compactification. These 5-branes all have 
topology $\R^4 \times \C$, where in the Calabi-Yau $uv + P(x,y)$ they
are located at specific points $x_i$ of the curve $P(x,y)=0$ along the
line $v=0$ (so they are parametrized by the remaining coordinate
$u$).   Having this extra pole for $y$ means that the Riemann surface
has extra tubes attached to it at $x=x_i$.

If we T-dualize this geometry to replace the Calabi-Yau $X$ by an NS5-brane
wrapping $\R^4 \times \Sigma$, the D5-branes, which are all transverse
to $\Sigma$, will become D4-branes. So we get an NS5-brane with a
bunch of D4-branes attached, that all end on the NS5-brane. This
configuration can be lifted to M-theory to give a single irreducible
M5-brane, now with ``spikes'' at the positions $x_1,\ldots,x_n$. So we
indeed see that the two kinds of fermions (or at least their sources)
are directly related and have the same effect on the geometry of the
Riemann surface.

\subsection{Counting BPS states and the topological string theory}

In Type II compactifications on Calabi-Yau geometries F-terms in the effective
action in four dimensions can be computed using topological string
theory techniques. We now want to see how these kind of computations
can be mapped to the I-brane model.

Starting point will be the end of the chain of dualities of section
\ref{ssec-chain}: the IIA compactification on $\R^3 \times S^1 \times
\wt X$ with $\wt X$ a (non-compact) Calabi-Yau geometry, where we wrap
a D6-brane along $S^1 \times \wt X$. This is the set-up of
Donaldson-Thomas theory. With the right background values of the
moduli turned on, the topological string partition function can be 
reproduced as an index that counts BPS states
degeneracies of D-branes in
this configuration . 

More precisely, the topological string theory partition function in
the A-model naturally splits in a classical and a quantum (or
instanton) contribution
\be
\label{Z-top}
Z_{top}(t,\la)  = \exp \left(-{t^3\over 6 \la^2} - {1\over 24} t \cdot
c_2\right) Z_{qu}(t,\la).
\ee
Here $t \in H^2(\wt X)$ is the complexified K\"ahler class, $\la$ the
topological string coupling contant, and $c_2=c_2(\wt X)$. As before
the wedge product is used to multiply forms. The quantum contribution is
decomposed as
$$
Z_{qu}(t,\la) = \sum_{g\geq 0} \la^{2g-2} \cF_g(t),
$$
with the genus $g$ free energy expressed in terms of the Gromov-Witten
invariants of degree $d$ as
$$
\cF_g(t) = \sum_d GW_g(d) e^{d \cdot t}.
$$

We can now use the fact that $Z_{qu}$ has a dual interpretation as the
Donaldson-Thomas partition function counting D0-D2-D6 bound states (we
ignore here a subtlety with the degree zero maps)
\cite{gsy,dvv,foam,mnop,ok-re-va}
\be
\label{DT}
Z_{qu}(t,\la) =  \sum_{n,d} DT(n,d) \, e^{-n\la} e^{d \cdot t}.
\ee
In this sum $n \in H_0(\wt X,\Z)\cong \Z$ and $d \in H_2(\wt X,\Z)$
are the numbers of D0-branes and D2-branes. The integers $DT(n,d)$
are the Donaldson-Thomas invariants of the ideal sheaves with these
characteristic classes. From the BPS counting perspective it is also
natural to add the exponential cubic prefactor in (\ref{Z-top}), since this
is nothing but the tension of a single D6-brane (including the
geometrically induced D2-brane charge).

In type IIA string theory set-up the complex parameters $\la$ and
$t$ can be expressed in terms of geometric moduli of the $S^1$ and the
Calabi-Yau $\wt X$, and the Wilson loops of the flat RR fields $C_1$
and $C_3$. In particular, we can write
\be
\label{lambda}
{\la} = {\beta \over \ell_s g_s} + i \theta, 
\ee
with
$$
\beta = 2\pi R_9 = \oint_{S^1} ds
$$ 
the length of the Euclidean time circle $S^1$, and $\theta$ the Wilson
loop
$$ \theta = \oint_{S^1} C_1.
$$ 
That is, $\lambda$ can be written as the holonomy of the complexified
one-form $ds/g_s+ i C_1$ (in string units). An important remark is
that, as expressed in equation (\ref{DT}), for BPS states $\theta$ and
$\beta$ only appear in the {\it holomorphic} combination
(\ref{lambda}). In the same way the parameter $t$ is given by the
integral of the complex 3-form $k \wedge ds/g_s+ i C_3$ over $S^1$.

It is rather trivial to also include the coupling to
D4-branes in this BPS sum. As explained before, such a bound state of $p$ D4-branes to
a D6-brane is given by a flux of the $U(1)$ gauge field on the
D6-brane.  We can think of this as a non-trivial first Chern class of
the line bundle over the D6-brane that wraps $\wt X$. Tensoring with
this extra line bundle will not change the BPS degeneracy, since the
moduli space of such twisted sheaves is isomorphic to that of the
untwisted one. The only thing that changes are the induced D0 and D2
charges, that shift as
\begin{eqnarray}
d & \to & d - {1\over 2}p^2 - {1\over 24}c_2\\
n & \to & n + d\cdot p + {1\over 6}p^3 + {1\over 24} p \wedge c_2.
\nonumber
\end{eqnarray}
So, if we also include a sum over the number of D4-branes, weighted by
a potential $v$, we get a generalized partition function 
\begin{eqnarray}\label{fermionpartfunc}
Z(v,t,\la) &=& \sum_{p \in H^2(\wt X,\Z)}e^{p(v-t^2/2\la)} e^{-(p^2/2+c_2/24)t}e^{-(p^3/6 + p
  c_2/24)\la}e^{-t^3/6\la^2}\ Z_{qu}(t+p\la,\la)\nonumber \\
 &=&  \sum_{p \in H^2(\wt X,\Z)} e^{pv}Z_{top}(t+p\la,\la).
\label{theta}
\end{eqnarray}
Apart from adding the D6-brane tension $-t^3/6\la^2$, we have also
added the tension $-p t^2/2\la$ of the D4-branes.  The structure
(\ref{theta}) will be the object that we want to identify with the
I-brane partition function and that should be computable in terms of
free fermions. It should be remarked that the partition function
(\ref{fermionpartfunc}) was also found in \cite{dvvonk}, where a dual
object was studied: a NS 5-brane wrapping $\wt X$. Also in
\cite{nekrasov-okounkov} a partition function of this type was
considered and directly related to fermionic expressions.

Let us now follow this D-brane set-up through the duality chain.
Before we do this, though, let us note that the above expression for
$Z(v,t,\la)$ has an interesting limit for $\la \to 0$, where only
genus zero and one contribute. In that case we have
\be
\label{pert}
 Z_{top}(t+p\la) \sim \exp\left[{1\over \la^2}\cF_0(t) + {1\over
  \la} p^I Z_I + {1\over 2} p^I p^J \tau_{IJ} + \cF_1(t) +
\cO(\la) \right].
\ee
If we now subtract the singular terms (which have a straightforward
interpretation as we shall see in a moment), we are left with the
familiar $\la=0$ answer
$$
Z(v,t) = \sum_p e^{{1\over 2} p\cdot \tau \cdot p + p \cdot v}
e^{\cF_1}.
$$

\subsection{Topological strings and I-branes}

If we lift the configuration of the previous section to M-theory, we
end up with a purely geometric compactification on $TN \times S^1
\times \wt X$. Asymptotically this geometry has a $T^2$ fibration over
$\R^3$.  The ratio of the radii of the two circles of this two-torus
is given by
$$
{\beta \over 2\pi g_s\ell_s}= {R_9 \over R_{11}}.
$$
The complex modulus of this $T^2$ is given by $\lambda/2\pi i$. If we
exchange the roles of the 9th and 11th dimension, we will perform a
modular transformation or S-duality
$$
\lambda \to 4\pi^2/\lambda.
$$ 
Equivalently, in the dual IIA compactification on $TN \times \wt X$
the asymptotic values for the radius of the circle fibration and
the graviphoton Wilson line are given by $-1/\lambda$. In the somewhat
singular limit $\beta \to 0$ we are left with $\lambda=\theta/2\pi$
and so after the duality we have a graviphoton proportional to
$1/\lambda$. Now this Wilson loop only makes sense asymptotically,
since this $S^1$ is contractible in $TN$. In fact, the graviphoton
gauge field is given by
$$
C_1 = {1 \over \lambda} \eta,
$$
where the one-form $\eta$ is our friend (\ref{eta}). This field
is not flat but has curvature
\be
\label{Gtwo}
G_2 =dC_1 = {1\over \lambda} \omega.  
\ee 
Here $\omega$ is the unique harmonic 2-form that is
invariant under the tri-holomorphic circle action (\ref{omega}). Note that this is a
natural choice, since in the case of $TN_1$ this form reduces to the
usual constant self-dual two-form in the center $\R^4$ (up to
hyper-K\"ahler rotations). 

Summarizing, the topological string partition function will be
reproduced by a IIA compactification on $TN \times \wt X$ with
graviphoton flux given by (\ref{Gtwo}). Note that in this case the
graviphoton is {\it inversely} proportional to the topological string
coupling!

It is now straightforward to follow this flux further through the
duality chain. In the type IIB compactification on $TN \times X$ the
self-dual 5-form RR field $G_5$ is given by
$$
G_5 = {1\over \la} \omega \wedge \Omega.
$$
with $\Omega$ the holomorphic $(3,0)$ form on the Calabi-Yau
$X$. 

We will now T-dualize this configuration to the IIA background that
includes a NS5-brane. In that case there is 4-form RR-flux
\be
\label{Gflux}
G_4 = {1 \over \la} \omega \wedge dx \wedge dy.
\ee
Here $dx \wedge dy$ is the $(2,0)$ form on the complex surface $\B$.
This 4-form flux can be directly lifted to $M$ theory, where we have
the geometry
$$
TN \times \cB \times \R^3.
$$

Now we have to discuss what the interpretation of this flux is, if we
reduce to IIA theory along the $S^1$ inside $TN$ to produce our system
of intersecting branes. In that case we will have an extra set of
D6-branes with geometry $\B \times \R^3$. We want to argue that the
$G_4$ flux becomes a constant NS $B$-field on their world-volumes.

As a preparation, let us recall again how the world-volume fields of
the D6-branes are related to the $TN$ geometry in the M-theory
compactification. First of all, the centers $\vec{x}_a$ of the $TN$
manifold are given by the vev's of the three scalar Higgs fields of
the 6+1 dimensional gauge theory on the D6-brane.  In a similar
fashion the $U(1)$ gauge fields $A_a$ on the D6-branes are obtained
from the 3-form $C_3$ field in M-theory. More precisely, if $\w_a$ are
the $k$ harmonic two-forms on $TN_k$ introduced in section 2, we have
a decomposition
$$
C_3 = \sum_a \w_a \wedge A_a.
$$
We recall that the forms $\w_a$ are localized around the centers
$\vec{x}_a$ of the $TN$ geometry. So in this fashion a bulk field gets
replaced by a brane field. This relation also holds for a single
D6-brane, because $\w$ is normalizable in $TN_1$.

As a direct consequence of this, the reduction of the 4-form field
strength $G=dC$ can be identified with the curvature of the gauge
field
$$
G_4 = \sum_a \w_a \wedge F_a.
$$
Combining this relation with the presence of the flux (\ref{Gflux}), we
find that in the I-brane configuration the D6-branes support a
constant flux
$$
\sum_a F_a = {1\ \over \la} dx \wedge dy.
$$
There is simple and equivalent way to induce such a constant magnetic
field on all of the D6-branes: turn on a NS $B$-field in the IIA
background. We can therefore conclude that in the presence of the
background flux (\ref{Gflux}) translates into a constant $B_{NS}$
field induced on the surface $\B$
$$
B_{NS} = {1\over \la} dx \wedge dy.
$$
In the next section we will discuss the full consequences of this.

\section{Topological strings and $\cD$-modules}  \label{sec-D-modules}

As we have seen in the previous section, ignoring all non-compact
directions, the essence of the intersecting brane configuration is a
D6-brane on the complex surface $\B$ with a D4-brane wrapped along the
curve
$$
\Sigma: \ \  P(x,y)=0.
$$
In addition a constant $B$-field
\be
\label{B}
B = {1\over \la} dx \wedge dy
\ee
is turned on. In this section we will argue that this set-up is most
naturally described using the formalism of $\cD$-modules.  But at this
point we first want to point out one immediate and more elementary
consequence of the $B$-field. The presence of the flux induces a
$U(1)$ gauge field on the D6-brane
\be
\label{A}
A = {1\over \la} y dx.
\ee
For the 4-6 strings we have to restrict $A$ to $\Sigma$. Therefore the
chiral fermions are coupled to a non-zero $U(1)$ gauge field. This
background gauge field gives a contribution to the effective action on
$\Sigma$ of the form
$$
\cF = {1\over 2} \sum_I \oint_{a^I}A \oint_{b_I}A.
$$
Here $(a^I,b_I)$ is a canonical basis of $H_1(\Sigma,\Z)$. Plugging
the expression for $A$ we obtain precisely the (genus zero)
prepotential of topological string theory
\be
\label{Fzero}
\cF = {1\over \la^2} \cF_0,
\ee
where, as always,
$$
Z^I = \oint_{a^I} ydx,\qquad \d_I \cF_0 = \oint_{b_I} ydx.
$$
In fact, we can also include a non-trivial flux $p^I$ through the
cycles $a^I$. This will give a second contribution to the free energy
given by
\be
p^I \oint_{b_I} A = {1\over \la}p^I \d_I \cF_0.
\label{p-F} 
\ee
We recognize the contributions (\ref{Fzero}) and (\ref{p-F}) as the
genus zero contributions in the expansion for small $\la$ of the
general expression (\ref{pert}).

\subsection{I-branes and $\cD$-modules}

Let us now explain how $\cD$-modules naturally appear in the I-brane
set-up. (See also \cite{langlands} for a much more involved setting.)
First of all, by very general arguments the algebra $\cal A$ of open
string fields on the D6-brane is naturally non-commutative. This is a
consequence of the fact that the Riemann surface that describes the
interaction
$$
\cA \otimes \cA \to \cA
$$
only has a cyclic symmetry (see fig. 5). 

\begin{figure}[t]
\begin{center}
\includegraphics[width=10cm]{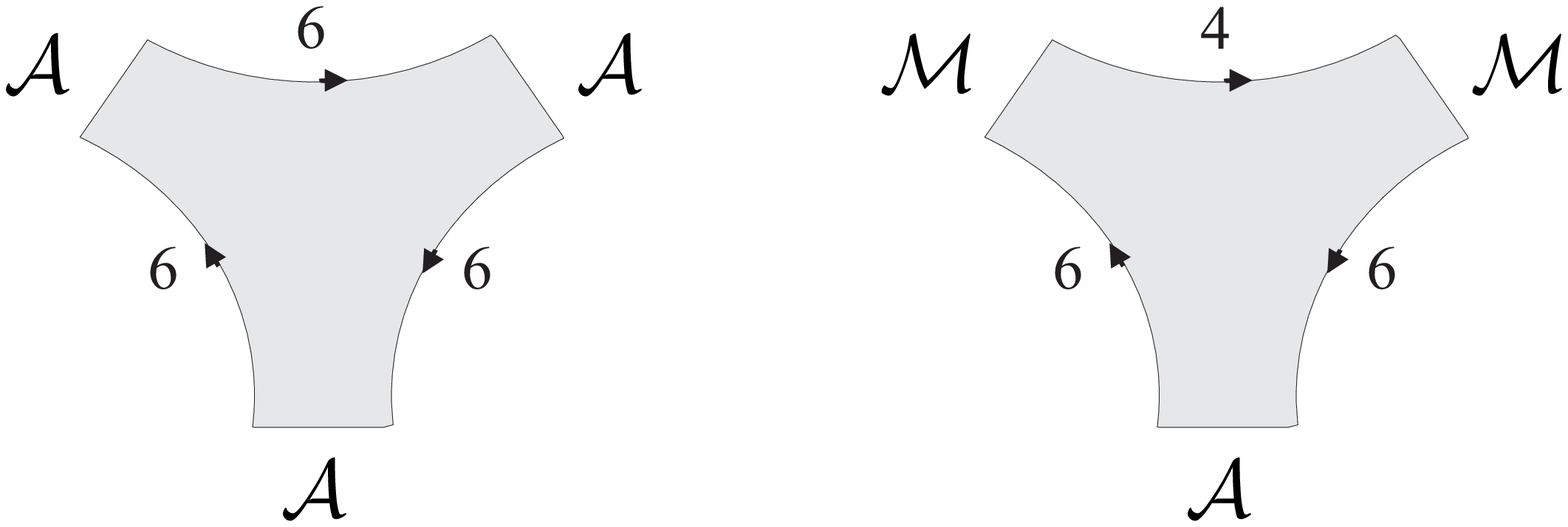}\\[5mm]
\parbox{12cm}{\small \bf Fig.\ 5: \it The 6-6 strings naturally form a
  non-commutative algebra $\cA$, whereas the 4-6 strings are a module
  $\cM$ for the algebra $\cA$.}
\end{center} 
\end{figure}

This non-commutativity is particularly clear if one includes a
$B$-field as in (\ref{B}). One simple way to see this is, that in the
presence of such a $B$-field a gauge field $A$ is induced that couples
to the open strings. The gauge field satisfies $dA=B$ and can be
chosen as
$$
A = {1\over \la} y dx.
$$
Therefore, on the 6-6 strings there is a boundary term
$$
\int A = \int dt\ {1\over \la} y \dot x.
$$
If we reduce this term to the zero-modes of the strings we get the
usual term of quantum mechanics, where $\la$ plays the role of
$\hbar$. Therefore the coordinates $x$ and $y$ become non-commutative
operators
$$
[\hat x,\hat y] = \la.
$$
So, the zero-slope limit of the algebra $\cA$ becomes the
Heisenberg algebra, generated by the variables $\hat x, \hat y$. This
is also known as the non-commutative or quantum plane. In the case
where $\cB = \C^2$ we can identify this algebra with the algebra of
differential operators on $\C$
$$
\cA \cong \cD_\C.
$$
The algebra $\cD_\C$ consists of all operators
$$
D = \sum_i a_i(x) \d^i,\qquad \d = {\d \over \d x}.
$$
Here we have identified $\hat y = - \la \d$.

Now suppose that we add some other brane to this system, in our case a
D4-brane localized on $\Sigma$. We pick $x$ as our local coordinate,
so that, once restricted to the curve, the variable $y$ is given by a
function $y=p(x)$. Now the space of 4-6 open strings, that we will
denote as $\cM$, is by definition a module for the algebra $\cA$ of
6-6 strings. This is a simple consequence of the fact that a 6-6
string acting on a 4-6 string produces again a 4-6 string, as depicted
in fig. 5. Therefore there is an action
$$
\cA \times \cM \to \cM.
$$
(More completely, $\cM$ is a $\cA$-$\cB$ bimodule, where $\cB$ is the
algebra of 4-4 open strings.)

Modules for the algebra of differential operators are called
$\cD$-modules. In this case we are interested in $\cD$-modules that in
the semi-classical limit reduce to curves or equivalently
Lagrangians. Such $\cD$-modules are called holonomic.

So we can draw the following conclusion: in the presence of 
a background flux, the chiral fermions on the I-brane should no
longer be regarded as {\it local fields} or sections of the spin bundle
$K^{1/2}$. Instead they should be interpreted as sections of a
non-commutative $\cD$-module. 

Notice that if $\Sigma$ is a non-compact curve, having marked points
at infinity, the symplectic form $dx \wedge dy$ becomes very large at
the asymptotic legs. This can be see by using the appropriate
variables at infinity $x'=1/x$ and $y'=1/y$. In these variables the
$B$-field becomes singular which means that the non-commutativity goes
to zero. This explains why it makes sense to speak about the usual
free chiral fermions at infinity, and to discuss their nontrivial
transformation properties from patch to patch as in \cite{adkmv}.
Considering compact spectral curves seems to be much more involved
from this perspective.

Let us explain such structures with a simple example of a
$\cD$-module. We start with the commuting case, in which the
algebra $\cA$ is given by the ring $\cO$ of functions on the plane. If
the spectral curve $\Sigma$ is given by $P=0$, then we can write $\cM
= \cO_\Sigma$ as the quotient
$$
\cM = \cO / {\cal I}_\Sigma,
$$
where ${\cal I}_\Sigma =\cO \cdot P$ is the ideal of functions vanishing on
$\Sigma$.

Now suppose that $P$ is not a polynomial, but a differential operator
$$
P \in \cD.
$$ 
Then we can similarly define a $\cD$-module as an equivalence class
of differential operators
$$
\cM = \cD/\cD P.
$$
One way to think about such a module is in terms of a formal solution
to the equation
\be
\label{P}
P \Psi =0.
\ee
Clearly, any differential operator of the form $D \cdot P$ will
annihilate $\Psi$, so that $\cM$ can also be realized as the vector
space of expressions of the form
$$
\cM = \{D\Psi;\ D\in \cD\}.
$$

Mathematically, the module $\cM$ captures the solutions of the
differential equation (\ref{P}) in the following fashion. Suppose that
one would want to solve this equation for a function $\Psi$ that takes value
in some function space $\cV$, {\it e.g.}\ the space of
square-integrable functions on $\R$. Such a space $\cV$ is itself a
$\cD$-module. Namely, $\hat x$ and $\hat y$ are realized as
multiplication by $x$ and the differential $-\la \d_x$. One of the
important properties of $\cD$-modules is that the space of solutions
of $P\Psi=0$ in $\cV$ is given by algebra homomorphisms
$$
{\rm Hom}_\cD(\cM, \E)
$$
where the gauge bundle $\E$ is in general some torsion-free sheaf on $\Sigma$.

The $\cD$-module $\cM$ and the corresponding differential operator $P$
should be considered as the non-commutative generalization of the
classical curve $\Sigma$. This is the ``quantum spectral curve'' from
the theory of quantization of integrable systems, as is known from the
geometric Langlands perspective \cite{langlands}.  Within the context
of string theory it is clear that there should be a {\it unique}
$\cD$-module that corresponds to the curve $\Sigma$. This prescription
should fix possible normal ordering ambiguities in $P$. It would be
interesting to understand this directly from the mathematical
formalism.

\subsection{A simple example: rational curves}

Let us illustrate this with a very simple example: the curve
$y=0$. Topologically $\Sigma=\C$ can be viewed as a disc and in the
fermion CFT it will correspond to a state in the Fock space $\cF$.  In
this case the corresponding $\cD$-module just consists of the
polynomials in $x$
$$
\cM = \C[x]
$$
and $\hat y$ is realized as $-\la \d_x$. The one-form $ydx$ vanishes
identically. The free fermion theory based on this module consists of
the usual vacuum state $|0\rangle$. 

But now we can make a small variation, by picking the curve
$$
y = p(x),
$$
with $p(x)$ some function.  In this case the (meromorphic) one-form is
$p(x)dx$. The corresponding $\cD$-module is still isomorphic to
$\C[x]$ as a vector space, but now the operator $\hat y$ is
represented as
$$
\hat y = -\la \d_x + p(x).
$$
Of course, there is an obvious map between these two modules: we
simply multiply the functions $\psi(x) \in \C[x]$ as
$$
\psi(x) \to e^{-S(x)/\la} \psi(x),\qquad \d S(x)=p(x)dx.
$$
In the quantum field theory, where $\psi(x)$ becomes an operator
acting on the Fock space $\cF$, this correspondence is represented by
a linear map $U$ such that
$$
U \cdot \psi(x) \cdot U^{-1} = e^{-S(x)/\la} \psi(x).
$$
If $S(x) = \sum_k t_k x^k$ such a map is given by
$$
U = \exp \sum_k {1\over \la} t_k \alpha_{k}
$$
where $\sum_k \alpha_k x^{-k-1} = \d\phi(x) = :\psi^\dagger \psi(x):$
is the usual mode expansion of the $U(1)$ current. Here we use that
$[\a_k,\psi(x)]= x^k \psi(x)$. The corresponding state in the Sato
Grassmannian is given by $U|0\rangle$. Since $\alpha_k |0\rangle=0$
for $k\geq 0$, this state is only different from the vacuum if the
function $S(x)$ (or $p(x)$) has poles.

\subsection{Conifold and c=1 string}

Another illustrative case is a canonical example in any text on
$\cD$-modules (see {\it e.g.} \cite{kashiwara,bernstein}), namely the
operator
$$
P = - \la x\d_x - \mu.
$$
Now there is an interesting singularity at $x=0$.

In the physical context this case corresponds to one of the best
studied examples in string theory: the so-called $c=1$
string (see {\it e.g.}  \cite{ghoshal-vafa, imbimbo-mukhi,
  gho-imbi-mukhi,mukhi-vafa, dijk-moore-pless,distler-vafa,
  kleb-gross}). At the self-dual radius the partiton function of the
$c=1$ string is given by
\bea
 \mathcal{F}_{c=1}(\mu,\la) 
 &=&  -{1 \over 2} \left( {\mu \over \la} \right)^2 \log \mu + {1
  \over 12} \log \mu \notag \\
  &&\qquad +~ \sum_{g \ge 2}
(-1)^{g-1} \frac{B_{2g}}{2g(2g-2)} \left( {\la \over \mu} \right)^{2g-2},  \label{c1}
\eea
with $\mu$ the cosmological constant, $\la$ the string coupling, and
$B_{2g}$ the Bernouilli numbers. This model has a dual interpretation
in terms of B-model topological strings on the deformed conifold
geometry \cite{ghoshal-vafa}
$$
uv + xy = \mu.
$$
Geometrically the $c=1$ model is thus characterized by the presence of
a holomorphic curve in $\C \times \C$ defined by 
\bea\label{conifold}
\Sigma:\ xy = \mu.
\eea
The curve $\Sigma \cong \C^*$ has two asymptotic regions, $U: x \to
\infty$ or $V: y \to \infty$.  Therefore in the usual operator
formalism there is a linear operator $S$ that acts between the two
state spaces associated to the two boundaries.

First consider the limit $\la=0$. In this regime the I-brane degrees
of freedom are just conventional chiral fermions $\psi$ on $\Sigma$,
which can be expanded as
\begin{align*} &\psi(x) =
\sum_{n \in \Z+1/2} \psi_n x^{-n - 1/2} \sqrt{dx} \quad \mbox{and}
\quad 
\psi^\dagger(x) = \sum_{n \in \Z+1/2} \psi_n^\dagger x^{-n - 1/2} \sqrt{dx}
\quad \mbox{in $U$}, \\
     & \wt\psi(y) =
\sum_{n \in \Z + 1/2} \widetilde{\psi}_n y^{-n - 1/2} \sqrt{dy} \quad \mbox{and}
\quad 
\wt\psi^\dagger(y) = \sum_{n \in \Z+1/2} \widetilde{\psi}_n^\dagger y^{-n - 1/2}
\sqrt{dy} \quad \mbox{in $V$}.
\end{align*}
The genus 1 part $\cF_1$ of the free energy is obtained from the
partition function of $\psi$. It can be computed by assigning the
usual Dirac vacuum $| 0 \rangle$ to $U$ and likewise the conjugate
state $\langle 0 |$ to $V$. The partition function is then computed as
$$
Z = \langle 0 |S|0 \rangle,
$$ 
where $S$ transforms $\psi$ from the $V$-patch to the $U$-patch, {\it
  i.e.} $\wt\psi(y)= S \psi(x) S^{-1}$. Substituting $y=\mu/x$ gives
$S \psi_n S^{-1} = {\mu^{n}}~ \widetilde{\psi}_{-n}$. Since the state
$|0\rangle$ can be thought as a semi-infinite wedge built up from the
differentials $dx^{n+1/2}$ and similarly $\langle 0|$ is made up from
the wedge product of $dy^{n+1/2}$ we find that
\begin{align}\label{eq:F1c=1}
e^{\cF_1} = \langle 0 |S|0 \rangle =  
\prod_{n\ge 1/2} \mu^{-n}. 
\end{align}
This expression can be computed using $\zeta$-function regularization
to give the familiar answer $\cF_1 = {1\over 12} \log \mu$.

In order to go beyond 1-loop, we should think in terms of
$\cD$-modules. The $c=1$ $\cD$-module is represented by
\begin{align*}
&\cM = \cD / \cD P, \quad \mbox{with} ~ P=-\la x \d_x - \mu ~
\mbox{in}~U, \\
&\widetilde{\cM} = \cD / \cD \widetilde{P}, \quad \mbox{with} ~
\widetilde{P} = \la y \d_y - \mu + \la ~ \mbox{in}~V.
\end{align*}
Nonetheless, it is easily seen that any solution $u(x)$ of
$Pu=0$ gives a solution $v=1/yu(y)$ of $\widetilde{P}v=0$ and vice
versa. Hence $\cM$ and $\widetilde{\cM}$ are equivalent.
Notice that the fundamental solution of $Pu=0$ is given by 
$\Psi = x^{-\mu/\la}$. Furthermore, since any element in
$\cM$ can be reduced to either $x^m$ or $y^m$ using the relation $xy =
\mu$, we can represent the action of $\cD$ on this module as 
\begin{align*}
& \hat{x} (x^m) = x^{m+1} \\ 
& \hat{y} (x^m) =  \left(- \la \d_x + \frac{\mu}{x}
\right)  x^{m}\\
& \hat{x} (y^m) =  \left( \la \d_y + \frac{\mu -
    \la}{y} \right) y^{m} \\ 
& \hat{y} (y^m) = y^{m+1}.  
\end{align*}

A basis of a representation of $\cM$ on which $\hat{x}$ and
$\hat{y}$ just act by multiplication by $x$ resp. differentiation
with respect to $x$ is given by
\begin{align*}
&v_{1m} (x) = x^m \cdot x^{-\mu/\la}, \\
&v_{2m}(x) =  \int dy~e^{-xy/\la} ~ y^{m-1} \cdot y^{\mu/\la}.  
\end{align*}
Indeed, differentiation with respect to $x$ clearly gives the same
result as applying $\hat{y}$. Moreover, multiplying $v_{2m}$ by
$x$ gives
\begin{align*}
x \cdot v_{2m} (x) &= \la \int dy~ e^{-xy/\la}
\frac{\partial}{\partial y} \left( y^{m-1+ \mu/\la} \right) = (\mu
+\la( m -1)) v_{2(m-1)}.
\end{align*}
Notice that if we had restricted ourselves to $\mathcal{O}_\Sigma$ and
taken the limit $\la \to 0$, we would have only found the standard basis
$x^m$, $m \ge 0$.  

Similarly, in the region $V$ one can verify that 
\begin{align*}
& \hat{y} (y^m) = y^{m+1} \\
& \hat{x} (y^m) =  \left(\la \d_x +
  \frac{\mu-\la}{x} \right)  x^{m}\\
& \hat{y} (x^m) =  \left( -\la \d_y  +
  \frac{\mu}{y} \right) y^{m} \\ 
& \hat{x} (x^m) = x^{m+1}.  
\end{align*}
Hence in the
representation of $\widetilde{\cM}$ defined by
\begin{align*}
&\tilde{v}_{1m} (y) = y^{m-1} \cdot y^{\mu/\la}, \\
&\tilde{v}_{2m}(y) =  \int dx~e^{xy/\la} ~ x^m \cdot x^{-\mu/\la},
\end{align*}
$y$ and $\partial_y$ act in the usual way. Since we moved
over to representations of the $\cD$-module where the differential
operator acts as we are used to, the $S$ transformation that connects
the $U$ and the $V$ patch and thereby exchanges $\hat{x}$ and
$\hat{y}$ must be a Fourier transformation. This is clear from the
expressions for the basis elements $v$ and $\tilde{v}$: $S$
interchanges $v_{1m}(x)$ with
$\tilde{v}_{2m}(y)$, and $v_{2m}(x)$ with $\tilde{v}_{1m}(y)$. So we
immediately find the result of \cite{adkmv}.

The partition function $Z$ of this system can now be easily computed in two
ways. First exactly, to give the all-genus answer. And secondly we use
the stationary phase approximation to show the relation with the
previous calculations. Notice that $v_{2m}(x)$ almost equals the gamma-function  $\Gamma(x) = \int_0^{\infty} dt~e^{-t}~t^{x-1}$. Indeed, 
\begin{align*}
v_{2m}(x) =  \frac{\la}{x} \int dy'~e^{-y'} ~ \left( \frac{\la y'}{x}
\right)^{m-1+\mu/\la} =  \left( \frac{\la }{x}
\right)^{m+\mu/\la} \Gamma(m+ \mu/\la).
\end{align*}   
Comparing this with $\tilde{v}_{1m}(y)$ implies that the free energy
$\cF$ equals 
\begin{align*}
\cF \left( \mu, \la \right) &=  \sum_{m \ge 0} \left(m+{\mu \over \la}\right) \log \left( \frac{\la }{\mu} \right) +  \log \Gamma(m+ \mu/\la).
\end{align*}
This function satisfies the recursion relation
\bea
\cF \left( {\mu \over \la} + {1 \over 2} \right) - \cF \left( {\mu \over \la} - {1 \over 2} \right) &=&  \left( {1 \over 2} - {\mu \over \la} \right) \log \left( {\la \over \mu} \right) - \log \Gamma \left( {\mu \over \la} - {1 \over 2} \right), \notag
\eea
from which we conclude (see Appendix A in \cite{nekrasov-okounkov}) that $\cF$ is the well-known answer for the $c=1$ string, up to linear terms in $\mu$ and $\la$.
\be
\cF_{c=1}(\mu,\la) ~=~ {1 \over 4} \int {dt \over t} ~ \frac{e^{-it\left( {\mu \over \la} \right)}}{\sinh^2(t/2)} + \mbox{Pol}_1(\mu,\lambda) \notag 
\ee
which indeed reproduces (\ref{c1}). In particular, we also 
recover $\cF_1$ from equation (\ref{eq:F1c=1}).

To reproduce the $\la$ expansion we can approximate the full genus answer
using the Euler-Maclaurin formula or apply the stationary phase
approximation to $v_{2m}$. This yields as
zeroeth order contribution to $v_{2m}$
\begin{align*} 
e^{-\mu/\la} \left( \frac{\mu}{x} \right)^{m-1+\mu/\la},
\end{align*}
while the subdominant contribution is given by 
\begin{align*} 
\sqrt{\frac{2 \pi \la \mu}{x^2}}.
\end{align*}
So in total we find that 
\begin{align*}
v_{2m}(x) = \sqrt{2 \pi \la}~(\mu/e)^{\mu/\la}~ x^{\mu/\la}~
\mu^{m-1/2}~ x^{-m}~ \psi_{qu}\left(\frac{\mu}{x}\right),
\end{align*}
which summarizes the contributions we found before, genus zero
$x^{\mu/\la}$ and genus one $\mu^{m-1/2}$, plus the higher
order contributions that are collected in $\psi_{qu}$.

\subsection{The topological vertex}

An important step to understand more general curves is the case of the
topological vertex \cite{vertex}. Its mirror is a genus zero curve with three punctures
given by the equation 
$$
x + y - 1 =0
$$
in $\C^* \times \C^*$. In this case the symplectic form is given by
$du \wedge dv$ where $u,v$ are logarithmic coordinates: $x=e^u$ and $y=
e^{v}$. The corresponding $\cD$-module is now given by the operator 
\cite{adkmv}
$$
P =  e^{u} + e^{-\la \d_u} - 1.
$$
$P$ is actually a difference operator, instead of a differential
operator, so we have to generalize the notion of a $\cD$-module
somewhat. This is a well-known procedure in the field of quantum
groups. These quantum groups appear because in the $\C^*$ case the
operators $\hat x$ and $\hat y$ now satisfy the Weyl algebra or
$q$-commutation relation
$$
\hat x \,\hat y = q \,\hat y \,\hat x,\qquad q=e^\la.
$$
The fundamental solution to $P\Psi=0$ is the quantum dilogarithm
$$
\Psi(u)= \prod_{n=1}^{\infty} {1 \over \left(1 - e^{u} q^{n} \right)}.
$$

The corresponding module $\cM$ for the Weyl algebra can again be
written in terms of the coordinate $u$ or in terms of the dual
variable $v$. There is another unitary map $U$ that implements this
transformation on the free fermion fields. Because of the hidden
cyclic symmetry of the vertex, this can be made transparant by
writing it as 
$$
e^{u_1} + e^{u_2} + e^{u_3} =0.
$$
Up to an overall rescaling of the three variables $u_i$, the map $U$
satisfies $U^3=1$.  This line of reasoning leads one directly to the
formalism of \cite{adkmv}, but we will not pursue this here in more
detail. We reach the important conclusion that the notion of a quantum curve, as
expressed in the concept of a (generalized) $\cD$-module, is the right
framework to derive the complicated transformations of
\cite{adkmv}. We will now use this correspondence in two concrete
examples of compact curves.

\subsection{Elliptic curves}

A well-studied example is the geometry mirror to the total space
of a rank two bundle over an elliptic curve
\be
\wt X:\ \mathcal{O}(-r)\oplus\mathcal{O}(r)\rightarrow T^2. 
\label{torus}
\ee
The latter has a description in toric geometry as glueing the toric
propagator to itself with a framing factor $r$. This changes the
intersection of $[T^2]$ with the 4-cycles that project onto $T^2$ into
$\pm r$ \cite{vafa-ym}.   In this section we show how one can
use the picture we have developed in this paper, and in particular
the free fermionic systems living on the boundary of the non-commutative
plane, to 
completely solve this model and
recover the existing results for the all genus topological string
amplitudes for this background.

This model has a simple interpretation in the B-model obtained after
mirror symmetry. Note that we can write the geometry (\ref{torus}) as
a global quotient of $\C^* \times \C \times \C$. If we pick toric
coordinates $(e^{u},e^{v},e^{w})$, the identification is
$$
(u, v,w) \sim (u + t, v + ru, w - ru).
$$
This transformation is an affine transformation consisting of a shift
$(t,0,0)$ and a linear map
$$
A = \left(\begin{array}{ccc}
1 & 0 & 0 \\
r & 1 & 0 \\
-r& 0 & 1 \\
\end{array}\right) \in SL(3,\Z).
$$
The linear transformation $A$ is the monodromy of the fiber, when we
view this non-compact CY as a $T^3$ fibration. Mirror symmetry will
now replace the torus fibers with their duals, and the monodromy $A$
with the dual monodromy $A^{-T}$. So the B-model can be described as a
quotient of the dual coordinates given by
$$
(u, v,w) \sim (u + t - rv + rw, v, w).
$$

In order to map this B-model to the NS 5-brane and finally the
I-brane, we have to perform one more T-duality on the combination
$v+w$. That coordinate is not touched by the action of the framing and
it will be subsequently ignored. If we relabel the coordinates as
$$
x=u,\qquad  y=v-w,
$$
we see that this gives indeed a $T^2$ curve, embedded as the zero
section $y=0$ in the geometry $\cal B$ defined as the quotient of $\C^*
\times \C$ by
\be
\label{toro}
(x,y) \sim (x + t - ry,y).
\ee

Now the A-model topological string partition function 
can be computed as \cite{vafa-ym, bryan-pandhari}
$$
Z_{top}(t,\la) = e^{-t^3/6 r^2 \la^2}Q^{-1/24} \sum_R Q^{|R|} q^{r \kappa_R/2},
$$
where $Q=e^{-t}$ with $t$ the K\"ahler parameter of the torus and
$q=e^{-\lambda}$, whereas $|R|$ is the number of boxes of the corresponding Young tableaux and $\kappa_R = 2 \sum_{\Box \in R} i(\Box)-j(\Box)$. After the mirror transformation $t$ becomes the
modulus of the elliptic curve $T^2$.  The instanton part of $Z_{top}$
can be rewritten in the form
\be
\label{T2}
Z_{qu}(t,\la) =  \oint \frac{dy}{2 \pi i y} \prod_{n=0}^{\infty}
\left(1+y\,Q^{n+1/2}q^{r(n+1/2)^2/2}\right)
\left(1+y^{-1}Q^{n+1/2} q^{-r(n+1/2)^2/2}\right)
\ee
which is familiar from \cite{dijkgraaf-mirror,douglas-2dym} in the case $r=1$.  In this model the genus zero answer does not have instanton contributions and
so is given entirely by the classical cubic form $\cF_0(t) = -{1\over
  6r^2}t^3$, while at genus one the classical and quantum
contributions combine into 
$$ \cF_1(t) = -\log \eta(Q).
$$
The $g$-loop contributions $\cF_g$, for $g >1$ and $r>0$, incorporate
only quantum effects.

In fact, it is well-known that this answer is reproduced by a chiral
fermion field with action \cite{dijkgraaf-chiral}
\be
\label{fermform}
S = \frac{1}{\pi} \int_{T^2} d^2x  \ \psi^\dagger\left(\bar{\partial} -
r\lambda \partial^2\right) \psi. 
\ee
We will rederive this same answer from the fermionic perspective
we have developed in this paper below.  For now note that
this action can be bosonized into \cite{douglas-2dym,gross-klebanov-bosonize}
\be
\label{T2boson}
S = \frac{1}{\pi} \int_{T^2} d^2x \left( \frac{1}{2} \partial \phi
\bar{\partial} \phi  -
\frac{r \lambda}{6} (\partial \phi)^3 \right), 
\ee
which is closely related to the Kodaira-Spencer field theory on the
Calabi-Yau manifold.  This Kodaira-Spencer theory reduces to a free
boson $\phi$ on a cylinder, while the framing quantizes into an action
of the zero mode of the $W^3$ operator \cite{adkmv}
$$
W^3_0 = \oint dx \frac{(\partial \phi)^3}{3}.
$$
This implies that $W^3_0$ defines how to glue the torus
quantum mechanically,
$$
Z_{top}= \Tr \exp \left(-{r \la \over 2}~ W^3_0 \right),
$$
explaining (\ref{T2boson}). The action of $W^3_0$ is quadratic in the
fermions and therefore acts on the single fermion states. 

The topological string partition function (\ref{T2}) is obtained as the
fermion number zero sector. Including also a sum over the $U(1)$ flux
gives the full fermion partition function that corresponds to the
I-brane. This can be thought of as a 
generalized Jacobi triple formula \cite{kanekozagier}. Adding the
classical contributions we obtain
\begin{eqnarray}
Z(v,t,\la) & = & e^{-t^3/6r^2\la^2} Q^{-1/24}\prod_{n=0}^{\infty} (1+yQ^{n+1/2}q^{r(n+1/2)^2/2})(1+y^{-1}Q^{n+1/2} q^{-r(n+1/2)^2/2}) \nonumber \\
& = &\sum_{p=-\infty}^{\infty} y^p e^{-t^3/6r^2\la^2} e^{-pt^2/2 r \la} Q^{p^2/2-1/24} q^{rp^3/6-rp/24} 
  Z_{qu}(t+rp\la,\la) \nonumber \\
& = & \sum_{p=-\infty}^{\infty} y^p Z_{top}(t+rp\la,\la).\nonumber
\end{eqnarray}
In the second line we have extracted a factor $e^{-t^2/2 r \la}$ out of $y$. This is the result of turning on flux in the I-brane set-up, and corresponds to the D4-brane tension on the BPS side. Notice that the combination $rp \in r \Z$. This is because $rp$ is the Poincar\'e dual of the four-cycle having intersection number $\pm r$ with $[T^2]$ . 
So this reproduces indeed formula (\ref{theta}) with an appropriate
choice of cubic form.  For $r=0$ this result reduces to the standard
Jacobi triple formula
$$
Z_{r=0}= {\theta_3(y,Q) \over \eta(Q)} = \sum_{n\in\Z} {Q^{n^2/2} y^n
  \over \eta(Q)}.
$$

We now come to deriving (\ref{fermform}) from the perspective of this
paper.  From the considerations of this paper it is clear that we have
a free fermion system living on $T^2$ with the {\it standard} action.
The only subtley has to do with the fact that $T^2$ is at the boundary
of a non-commutative plane and as we will see this is crucial in
recovering (\ref{fermform}).  From (\ref{toro}) we see that $x\sim
x+t-ry$.  If we treated $y$ as commuting with $x$ we could set it to
$y=0$ and we have a copy of the torus.  But here we know that $y$ does
not commute with $x$.  So we have a free fermion on a torus where the
modulus is changed from
$$t \rightarrow t-ry.$$
The variation of $t$ can be absorbed into the fermionic action by the 
usual
Beltrami differential $\mu_{\bar z}^z =\delta t$:

$$
S = \frac{1}{\pi} \int_{T^2} d^2x  \ \psi^\dagger\left(\bar{\partial} +
\mu \partial\right) \psi.
$$ 
Here we need to substitute $\mu =\delta t = -r y$.  In the classical
case where $y$ is commuting, this would give $\mu =0$ and we get the
same system as the usual fermions.  However, since $x$ and $y$ do not
commute, we should view $y=\lambda \partial_x$ leading to $\mu = -ry = -r
\lambda \partial_x$.  Substituing this operator for $\mu$ in the above
action reproduces (\ref{fermform}).  We have thus rederived the known
result for the topological string in this background from our
framework.

\subsection{Genus two curves}

An interesting generalization of the elliptic curve example is given
by a B-model geometry containing a genus two curve. This model can be
constructed using the topological vertex technology. Although the
vertex technology is able to deal with arbitrary curves, it might be
instructive to see this explicit case in more detail.

Let us start in the A-model with the toric diagram of the resolved
conifold $\cO(-1) \oplus \cO(-1) \to \P^1$ and identify the two pairs
of parallel external legs, as shown in fig. 6. We refer to this
geometry as $\tilde{X}$. The B-model geometry corresponding to
$\tilde{X}$ is a locally elliptic Calabi-Yau $X$, described by an
equation of the form $uv = P(x,y)$, where $P$ vanishes on some compact
genus two Riemann surface $\Sigma$.

\begin{figure}[h!]
\begin{center}
  \includegraphics[scale=.4]{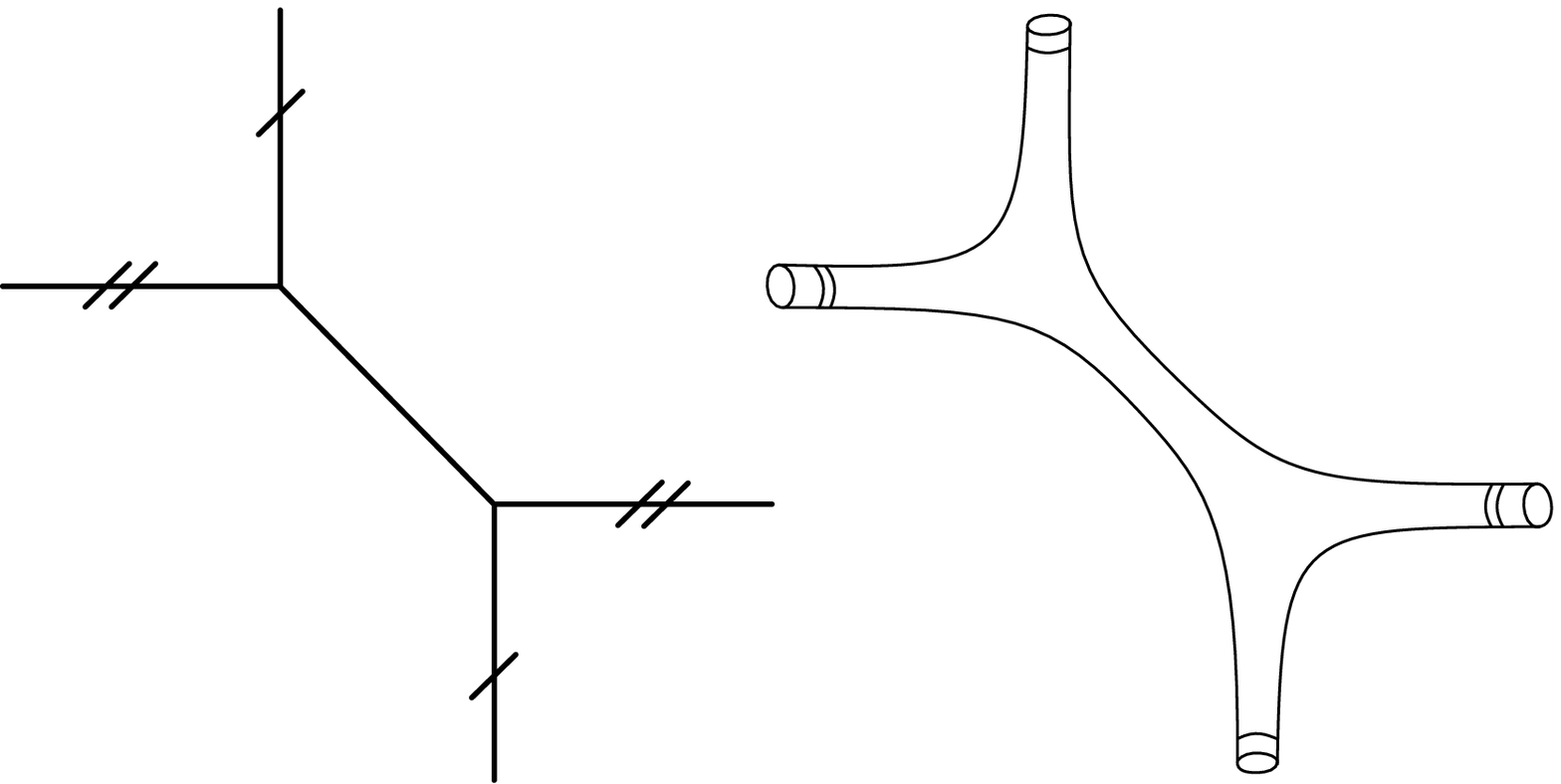}
\parbox{13cm}{\small \bf {Fig.\ 6:} \it The resolved conifold with identified legs (left) and its mirror (right).}
  \label{fig:genustwo}  
\end{center}
\end{figure}

This B-model geometry is well-studied in \cite{h-i-v} as an example of an 
elliptic threefold geometrically engineering a six-dimensional gauge theory 
on $\R^4 \times T^2$. The prepotential of this gauge theory is computed  
as the A-model partition function of $\tilde{X}$. Since this is a 
topological vertex calculation, the all-genus partition function is known. 
Moreover, instanton calculus in the six-dimensional gauge theory shows 
that it can be elegantly rewritten in terms of the equivariant elliptic
genus of an instanton moduli space. The equivariant parameter 
$q$ equals $e^{-\lambda}$ on the A-model side. 

Explicitly, the A-model on $\tilde{X}$ can be expressed in topological 
vertices as
\begin{align*}
Z_{qu}(Q_1,Q_2,Q_3) = \sum_{R_1,R_2,R_3} Q_1^{R_1} Q_2^{R_2} 
Q_3^{R_3} (-)^{l_{R_1}+l_{R_2}+l_{R_3}}C_{R_1 R_2 R_3} C_{R_1^t R_2^t R_3^t}, 
\end{align*}
where $Q_i$ represent the exponentiated K\"ahler classes of the legs
with attached $U(N)$ representations $R_i$, and where $C_{R_1 R_2
R_3}$ is the topological vertex. Notice that $C_{R_1 R_2 R_3}$ is
symmetric under permutations of the $R_i$, while in terms of the toric 
graph it is more natural to use the variables
\begin{align*}
Q_{\sigma} := Q_2 Q_3, \qquad Q_{\rho} := Q_1 Q_3, \qquad 
Q_{\nu} = Q_3,
\end{align*}
that exhibit the $\mathbb{Z}_2$ symmetry between $Q_{\sigma}$ and 
$Q_{\rho}$. Using these definitions
\begin{align*}
& Z_{qu}(q,\rho,\sigma,\nu) =  \sum_R
Q_{\rho}^{l_R} \prod_{\Box \in R} \frac{(1-Q_{\nu}
  q^{h(\Box)})(1-Q_{\nu}^{-1} q^{h(\Box)})} {(1-q^{h(\Box)})^2} \\
 & \ \ \times \prod_{k=1}^{\infty} \frac{(1-Q_{\sigma}^k Q_{\nu} q^{h(\Box)})
  (1-Q_{\sigma}^k Q_{\nu} q^{-h(\Box)}) (1-Q_{\sigma}^k Q_{\nu}^{-1}
  q^{h(\Box)}) (1-Q_{\sigma}^k Q_{\nu}^{-1} q^{-h(\Box)}) }
   {(1-Q_{\sigma}^k q^{h(\Box)})^2 (1-Q_{\sigma}^k q^{-h(\Box)})^2
     (1-Q_{\sigma}^k)}.  \notag
\end{align*}
And this may be rewritten as  \cite{li-2004}
\begin{align}\label{eq:prodAmodel}
&Z_{qu}(q,\rho,\sigma,\nu)= \sum_{k \ge 0} Q_{\rho}^k \chi
((\mathbb{C}^2)^{[k]};Q_{\sigma},Q_{\nu})(q,q^{-1})  \\
&= \prod_{j \in (\mathbb{Z}/2)_{\ge 0}}
\prod_{\scriptsize\substack{a,k \ge 0,\\ c,l \ge 1}} 
\prod_{b=-j}^{j}  \left(
\frac{(1-Q_{\rho}^l Q_{\sigma}^a Q_{\nu}^{c-1} q^{2b+k+1})
(1-Q_{\rho}^l Q_{\sigma}^a Q_{\nu}^{c+1} q^{2b+k+1})}
{(1-Q_{\rho}^l Q_{\sigma}^a Q_{\nu}^{c} q^{2b+k+2})
(1-Q_{\rho}^l Q_{\sigma}^a Q_{\nu}^{c} q^{2b+k})}
\right)^{(k+1)C(la,j,c)} \notag
\end{align}
with $b=-j, -j+1, \ldots,j-1,j$ and $q=e^{-\lambda}$, whereas the
coefficients $C(a,j,c)$ are related to the equivariant elliptic genus
of $\C^2$ in the following way
\begin{align*}
\chi(\C^2,y,p,q) =& \prod_{n \ge 1}
  \frac{(1-yp^nq)(1-y^{-1}p^nq^{-1})(1-yp^nq^{-1})(1-y^{-1}p^nq)}
{(1-p^nq)(1-p^nq^{-1})(1-p^nq^{-1})(1-p^nq)} \notag \\
=~& \sum_{a\ge 0}\sum_{j \in (\mathbb{Z}/2)_{\ge 0}, c\in \mathbb{Z}}
  C(a,j,c) p^a (q^{2j} + q^{2(j-1)} + \ldots + q^{-2j}) y^c. 
\end{align*}
Starting with the IIA background 
$TN_1 \times \tilde{X}$ and going backwards through the duality chain, we find 
ourselves in the I-brane set-up   
on $\R^3 \times T^4 \times \R^2 \times S^1$. The genus two curve $\Sigma$ is 
holomorphically embedded in the abelian surface $T^4$ by the Abel-Jacobi map. 
The I-brane is the intersection of a D4-brane wrapping $\R^3 \times \Sigma$
and a D6-brane wrapping $T^4 \times \R^2 \times S^1$. The aim of this
section is to give an interpretation of the above A-model
result on $\tilde{X}$ in the I-brane picture.

\subsub{The case $\lambda=0$}

As a result of the duality chain, we expect that the 1-loop free energy 
$\cF_{1,top}$ of the topological A-model yields the free energy 
$\cF_{1,boson} = - {1\over 2} \log \det \Delta_\Sigma$
of a free chiral boson on $\Sigma$. Another sum over the lattice of momenta 
should then result in the chiral fermion determinant.  
Since not only the A-model partition function, but also the partition 
function of chiral bosons on a genus two surface is known, we can perform 
an explicit check of these conjectures.  

The genus two curve $\Sigma$ can be given explicitely as the zero locus of a
generalized theta-function on $T^4$ \cite{h-i-v}, and its $2 \times 2$ period 
matrix is expressed in terms of the mirror K\"ahler moduli 
$\{\rho,\sigma,\nu\}$. Twenty-four chiral bosons on $\Sigma$ are described by 
$\Phi_{10}(\rho,\sigma,\nu)$, the unique automorphic form of weight 10 of 
$Sp(2,\mathbb{Z})$ \cite{mooretwoloop,bkmptwoloop}. The partition function of a single chiral boson may be written as the generating function of the elliptic genus of a 
symmetric product of $TN_1$'s \cite{Dijkgraaf:1996it}
\begin{align}
\label{eq:genustwo}
  Z_{boson}(\rho,\sigma,\nu)
& = \, \frac{1}{\Phi_{10}(\rho,\sigma,\nu)^{1/24}} \,=\, 
\sum_N e^{2 \pi i N \sigma} \chi_{\rho,\nu} \left(TN_1^N/S_N
\right) \nonumber \\
& = e^{-\pi i (\rho + \sigma+v)/12} \!\!
\prod_{(k,l,m)>0} \!\! \left( 1-e^{2 \pi i (k\rho + l\sigma +
      m\nu)} \right)^{-c(4kl-m^2)}.
\end{align}
The numbers $c(4kl-m^2) \in \Q$ are defined in terms of the elliptic genus of 
$K3$. 
\begin{align*}
\chi(K3,\tau,z) = \sum_{h \ge 0, m \in \mathbb{Z}} 24 \, c(4h-m^2) e^{2 \pi
  i(h\tau+mz)}, 
\end{align*}
the unique weak Jacobi form of index 1 and weight 0. Here we have used
that $K3$ is generically a combination of 24 $TN_1$'s.

Looking back at the A-model partition function (\ref{eq:prodAmodel}),
it turns out that singling out the $\lambda^0$-part yields a sum
of the same form as (\ref{eq:genustwo})
\begin{align*}
\cF_{1,qu}(\rho,\sigma,\nu)= \tilde{c}(kl,m)
\prod_{(k,l,m)} \log \left( 1-e^{2 \pi i (k\rho + l\sigma +
      m\nu)} \right),
\end{align*}    
where the coefficients $\tilde{c}(kl,m)$ are related to the Fourier coefficients $C(a,j,c)$ as
\begin{align*}
&\tilde{c}(kl,m) = \\
- \sum_{j \in (\mathbb{Z}/2)_{\ge 0}} \sum_{b=-j}^{j}  
\Big[ \left( 2b^2-\frac{1}{12} \right) &\left(
  C(kl,j,m+1) + 
 C(kl,j,m-1) \right) 
- \left(4b^2 + \frac{5}{6} \right) C(kl,j,m) \Big].
\end{align*}

Actually, precisely the same relation can be found by relating the
elliptic genus of $K3$ and the equivariant elliptic genus of $\C^2$:
\begin{align}\label{eq:ellgenK3}
\chi(K3,y,p) &= - y^{-1} \int_{K3} x^2 \prod_{n\ge 1} 
\frac{(1-yp^{n-1}q^{-1})(1-yp^{n-1}q)(1-y^{-1}p^{n}q^{-1})
(1-y^{-1}p^{n}q)}
{(1-p^{n-1}q^{-1})(1-p^{n-1}q)(1-p^{n}q^{-1}) (1-p^{n}q)} \notag \\
&= - \int_{K3} x^2 \left( \frac{y+y^{-1}-q-q^{-1}}{A(x)A(-x)} \right) \chi(\C^2,y,p,q),
\end{align}
with $q = e^x$ and $A(x) := \sum_{k\ge0}\frac{x^k}{(k+1)!}$. So after adding the
classical contributions to $\cF_{1,qu}(\tilde{X})$ (proportional to
the K\"ahler class $t =\rho+ \sigma + \nu$) we may conclude that the
chiral boson determinant on $\Sigma$ equals the 1-loop partition
function of the B-model topological string on $X$ 
\begin{align*}
Z_{boson}(\rho,\sigma,\nu)=e^{\cF_{1,top}}(\rho,\sigma,\nu).
\end{align*}

In order to find the total contribution for $\lambda$ small, we have
to consider $\cF_{0,top}$ as well. In the B-model on $X$ its
second derivative has a simple interpretation: it's just the period
matrix $\tau_{ij}$ of the genus two curve $\Sigma$. In terms of the mirror map,
these periods will have classical contributions linear in $\rho$,
$\sigma$ and $\tau$, and quantum corrections determined by $Z_{qu}(\tilde{X})$. 
We will write these down in the next paragraph. Right now, let us
conclude with 
\begin{align*}
Z_{fermion}(\rho,\sigma,\nu)=\sum_{p_1,p_2 \in \Z} e^{\pi i p_i
  \tau_{ij} p_j} e^{\cF_{1,top}}(\rho,\sigma,\nu).
\end{align*}

\subsub{Automorphic properties}

Knowing the full instanton partition function (\ref{eq:prodAmodel})
makes it possible to examine the $\lambda$-corrections to $\cF_{1,top}$
explicitly. In fact, let us start more generally with the
Gopakumar-Vafa partition function 
\begin{align*}
Z_{qu} = \prod_{\Sigma \in H_2} \prod_{j \in \Z/2} \prod_{b=-j}^j
\prod_{k=1}^{\infty} (1-q^{2b+k} Q^{\Sigma})^{(-1)^{2j+1}k N^{2j}_{\Sigma}}.
\end{align*}
In order to get the $g$-loop free energies we note that
\begin{align}\label{eq:expanding}
&\log \prod_{m \ge 1} (1- Y q^{m+l})^m =\notag \\
 &\quad = - \frac{1}{\lambda^2}\mbox{Li}_3(Y) + {\frac{1}{2}(l^2 -
  \frac{1}{6})} \log(1-Y) - \lambda^2 \left( \frac{1}{240} - 
\frac{l^2}{24} + \frac{l^4}{24} \right) \mbox{Li}_{-1}(Y)
 - \ldots \notag \\
&\quad =: - \sum_{g\ge 0} \lambda^{2g-2} P_{2g}(l) \sum_{n \ge 1} n^{2g-3} Y^n,  \notag
\end{align}
where the degree $2g$ polynomials $P_{2g}(l)$ are defined through the last
equality. Hence
\begin{align*}
\cF_{qu} = - \sum_{g \ge 0}  \lambda^{2g-2} \sum_{\Sigma \in H_2} \sum_{j \in \Z/2} \sum_{b=-j}^j
 (-1)^{2j+1}  P_{2g}(2b) N^{2j}_{\Sigma} \sum_{n=1}^{\infty} n^{2g-3}
 \left(Q^{\Sigma} \right)^n.
\end{align*}
Making this expansion for the genus two Calabi-Yau $X$ reveals that
the coefficients $$ c^g_{\Sigma} = \sum_{j \in \Z/2}\sum_{b=-j}^j
(-1)^{2j+1} P_{2g}(2b) N^{2j}_{\Sigma} $$ are the Fourier coefficients
of Jacobi forms $J_g(q,y) = \sum_{k,l} c_g(k,l) q^k y^l$ of weight
$2g-2$ and index 1. More precisely, we can write the $\cF_{g}$'s as
\begin{align*} 
\cF_{g}(\lambda;Q_{\rho},Q_{\sigma},Q_{\nu}) &= - \lambda^{2g-2}
\sum_{k,l,m} c_g(kl,m) \sum_{n\ge 1} n^{2g-3} (Q_{\rho}^k
Q_{\sigma}^l Q_{\nu}^m)^n \\
&=- \sum_{N \ge 0} Q_{\rho}^N \sum_{kn =N} n^{2g-3}
\sum_{l \ge 0, m} c_g(kl,m) Q_{\sigma}^{ln} Q_{\nu}^{mn}  \\
&= - \sum_{kn=N} N^{2g-3} \sum_{b=0, \ldots, k-1} k^{2-2g} \, Q_{\rho}^N~J_g
\left(\frac{n \sigma+b}{k},n \nu \right) \\
&= - \sum_{N \ge 0} Q_{\rho}^N T_{g,N} (J_g),
\end{align*}
where $T_{g,N}$ are Hecke operators acting on Jacobi forms of weight
$2g-2$. This implies that all $\cF_{g,top}$'s are lifts of Jacobi forms, and
therefore automorphic forms of $O(3,2,\Z)=Sp(4,\Z)$
\cite{borcherds95autforms}.

\subsub{Interpretation in the duality chain}

First of all, notice that the partition function of $\tilde{X}$ can be
build out of topological vertices, and as such is known to have an
interpretation in terms of chiral bosons and fermions
\cite{vertex,adkmv,eynard-orantin,bo-kl-ma-pa}. The duality chain
elucidates these  
observations: the chiral fermions can be
identified with the intersecting brane fermions. Moreover, the $B$-field
on the D6-brane makes it necessary to treat these fermions as
noncommutative objects, which gives an explanation for the nontrivial  
transformation properties in \cite{adkmv}.

In terms of the gauge theory picture we can just refer to 
\cite{h-i-v}. Here it is shown that the six dimensional gauge theory on $TN_1
\times T^2$ can be engineered with matrix model techniques, revealing
$\Sigma$ as the Seiberg-Witten curve, whose period matrix equals the
second derivative of $\cF_{0,top}$. 

Finally, the automorphic properties of $\cF_{top}$ seems to fit in
best in the M5-brane frame of the duality chain. Recall that 
S-duality relates IIB on $TN_1 \times X$ to a NS5-brane wrapping
around $TN_1 \times \Sigma$ in the background $TN_1 \times T^4 \times
\R^2$. This lifts to a M5-brane in M-theory on
$TN_1 \times T^4 \times \R^2 \times S^1$. Since the M5-brane partition
function is expected to be an automorphic form of $O(3,2,\Z)$
\cite{Dijkgraaf:1998xr}, this perspective offers a physical reason for
the automorpic properties.
 
Actually, we know exactly which Jacobi forms enter: $J_0 =
\phi_{-2,1}$ is the unique Jacobi form of weight $-2$ and index 1,
$J_1 = - \frac{1}{12} \phi_{0,1} = - \frac{1}{24} \chi(K3,q,y)$ as we
encountered before, $J_2 = \frac{1}{240} E_4 \phi_{-2,1}$, $J_3 =
-\frac{1}{6048} E_6 \phi_{-2,1}$ and $J_4 =\frac{1}{172800} E_4^2
\phi_{-2,1}$ etc. Interestingly, these can all be defined as twisted
elliptic genera of $TN_1$ in the sense that (compare with
(\ref{eq:ellgenK3})) 
\begin{align*}
J_g(q,y) = -y^{-1} \int_{TN_1} x^{4-2g} \prod_{n\ge 1} 
\frac{(1-yp^{n-1}q^{-1})(1-yp^{n-1}q)(1-y^{-1}p^{n}q^{-1})
(1-y^{-1}p^{n}q)}
{(1-p^{n-1}q^{-1})(1-p^{n-1}q)(1-p^{n}q^{-1}) (1-p^{n}q)},
\end{align*}
coinciding with the M5-brane point of view and longing for a two dimensional
conformal field theory interpretation.

\section{Summary and outlook}

In this paper we have provided a unifying point of view on various
four-dimensional supersymmetric gauge theories, by relating them to
two-dimensional conformal field theories and free fermion systems.
This is accomplished by realizing the supersymmetric gauge theories 
as a system of D4 and NS5-branes, which can be lifted to an M5-brane
configuration. By reducing this M5-brane to the so-called I-brane,\
i.e.~a system of D4 and D6-branes intersecting along a spectral curve
$\Sigma$, the free fermions arise as massless states of open strings
living on $\Sigma$. This curve plays a fundamental role in our
considerations, and in particular it encodes much information about
gauge theories. First of all, it relates to a number of
supersymmetries in gauge theories: $\mathcal{N}=4$ theories arise for
$\Sigma=T^2$, while $\mathcal{N}=2$ theories for curved $\Sigma$. 

If the number $k$ of D6-branes in the I-brane configuration is larger
than one, the world-volume of the corresponding gauge theory is related to a
non-trivial ALE space of $A_{k-1}$ type. This fact proved very useful 
while re-examining the connection between $\mathcal{N}=4$
theories and conformal field theories, originally discovered by
Nakajima \cite{nakajima} and further explained by Vafa and Witten as a
consequence of the S-duality \cite{vafa-witten}. In particular, we
showed that the full I-brane partition function is given simply by the
fermionic character, which reduces to the Nakajima-Vafa-Witten results
upon decoupling in the gauge theory limit. This perspective also
allowed us to derive the McKay correspondence from a string-theory
perspective. 

In the analysis of $\mathcal{N}=2$ theories from the I-brane viewpoint 
it was crucial to realize that the curve $\Sigma$ becomes non-commutative,
with coordinates becoming operators satisfying
$$
[\hat x,\hat y] = \la,
$$
while the non-commutativity parameter $\la$ gets identified with 
a value of NS $B$-field turned on along the D6-brane.
Starting from the I-brane with a single D6-brane 
we presented a duality chain which relates it to the B-model topological strings on a
local elliptic Calabi-Yau defined by an equation of the form 
$$
uv + P(x,y)=0.
$$ 
Simultaneously, the locus $P(x,y)=0$ represents the I-brane curve $\Sigma$ 
embedded in a non-commutative plane. This enables an interpretation of
the all-genus topological string partition function in terms of physical
non-commutative free fermions which only have support on $\Sigma$,
and, mathematically, leads us to their formulation in terms 
of holonomic $\cD$-modules. The relation to $\cD$-modules provides 
an elegant explanation of unusual transformation properties
of those fermions resembling Fourier transformations, and allows them
to be identified with the fermions introduced in \cite{adkmv}, 
thereby realizing the latter ones from a physical string-theoretic
point of view.  
$$
*\quad *\quad *
$$

There are various open problems related to the results described above, 
which we hope to address in the future. Let us name just a few of them.
First of all, we only considered gauge theories with extended
supersymmetry. However it would be interesting to connect also
$\mathcal{N}=1$ theories to I-brane configurations. This should be
possible, especially in view of the fact that relations between
$\mathcal{N}=1$ theories and topological strings are already
well-understood \cite{dijkgraaf-vafa}. Even more appealing would be to
understand metastable non-supersymmetric gauge
theories in terms of
I-branes~\cite{metastable-ISS,metastable-ABSV}. For example they can 
be realized in a system of D4-$\overline{\textrm{D4}}$ branes spanned
between NS5-branes \cite{metastable-MPS}, which could be implemented
into one duality frame arising in our considerations. 

Secondly, it would be interesting to extend the analysis of curved
I-branes to cases with more than just a single D6-brane. In
particular it might shed some light on a connection between
topological strings and Donaldson-Thomas invariants, so far understood
only in case of a single D6-brane wrapped on a Calabi-Yau manifold ---
which is the configuration ending the duality chain described
in section \ref{ssec-chain}. Understanding the physics of an arbitrary
number of wrapped D6-branes would extend known relations beteen  Donaldson-Thomas, Pandharipande-Thomas \cite{PT},
Gopakumar-Vafa and Gromov-Witten invariants and could give a deeper understanding of them. 

Furthermore, more general geometries on which gauge theories are
defined could be analyzed, together with corresponding I-brane
configurations. In this paper we focused mainly on $U(N)$ gauge
theories on Taub-NUT geometries, and explained how replacing them
by Atiyah-Hitchin spaces relates to a presence of orientifold planes
in the I-brane system and leads to $Sp/SO$ gauge groups. Those
$U/Sp/SO$ gauge groups are intimately related to various affine Lie
algebras, realized in terms of free fermions on the I-brane. However,
as shown in \cite{gno}, there is a long list of affine Lie algebras
which can be realized in terms of free fermions. Is it possible to
engineer I-brane configurations which would support all those kinds of
fermions and Lie groups? 

On the other hand, we might consider further examples of target space
curves, either using orbifold techniques and making contact with vast
literature on CHL orbifolds \cite{CHL-JS,CHL-DJS}, or by considering
toric curves of higher genus.  So far we have succeeded in writing
down the I-brane action for a propagator in a toric geometry, but not
yet for a three-vertex. It would be nice to be able to explain the
genus two example from such a perspective.

Finally, we have just made a first step in uncovering the relevance of
$\cD$-modules in string theory. Of course, these objects have already
entered the field through the Langlands program \cite{langlands},
where they relate to eigenbranes of the 't Hooft operator in the
four-dimensional twisted gauge theory. However, in that context the
$\cD$-modules appear as non-commutative structures on coisotropic
branes in the A-model, whereas our set-up is purely physical and
holomorphic.

Let us here make a few more remarks about the relation of
$\cD$-modules with integrable hierarchies. Recall that solutions of
the KP hierarchy can be written down in the form of elements $W = w_1
\wedge w_2 \wedge \cdots$ of a Grassmannian. They are characterized by
their index, which is determined by the projection of the subspace
onto the vacuum. The big cell is a dense subset of the index zero
Grassmannian and contains those elements that project bijectively onto
the Dirac vacuum. A dense subset of the big cell admits a geometrical
description in terms of a torsion-free sheaf $\E$ on an algebraic
curve $\Sigma$, together with a trivialization of both $\Sigma$ and
$\E$ at some point $x_{\infty}$. This is called the Krichever
correspondence. Holomorphic sections of $\E$ on $\Sigma/x_{\infty}$
determine an element $W$ of the Grassmannian. Since these sections can
be interpreted as fermions $\psi$ living on $\Sigma$, $V$ is the
subspace that $\psi$ sweeps out in the Hilbert space $L^2(S^1)$. Their
index being zero means that there are as many excitations as gaps. The
tau-function of the hierarchy in these cases has the interpretation as
the fermion determinant det$\bar{\partial}_{\E}$.

Interestingly,
examples as Hermitean matrix models are part of the big cell, but not
of this dense subset. From our I-brane perspective we expect that, in
analogy to the case of
classical free fermions on a Riemann surface, the total topological
partition function should be given by some generalized determinant of
non-commutative fermions, reducing to the usual fermion determinant
when $\lambda=0$. It would be exciting to find such a generalization. 

Yet another remark is that it is well-known that the total big cell has an
interpretation in terms of a class of $\cD$-modules $\cM$ on a disk; by
localizing the $\cD$-module $\cM$ at $x_{\infty}$ we find an element $W$ of the
big cell. In the work of Ben-Zvi and Nevins  \cite{ben-zvi-nevins}
this picture is extended (with so-called $\cD$-lattices) and the
notion of a corresponding Lax operator $K$ (with the help of
micro-opers) to the whole Grassmannian. This led them to studying
$\cD$-bundles on curves and paved the way for a natural relation with
Calogero-Moser systems. However, the $\cD$-bundles they introduce can be seen as
non-commutative analogues of torsion-free sheaves on a ruled surface,
whose support covers all of this surface. This is unlike the
$\cD$-modules we encounter, which just have their support on the
spectral curve $\Sigma$. Still, this might be an interesting arena to
explore further.  

And lastly, it would be great to connect to recent developments by
Eynard, Marino and others (see e.g.
\cite{eynard-orantin,bo-kl-ma-pa}), who formulate the B-model on
locally elliptic Calabi-Yau's given by an equation of the form
$uv=P(x,y)$ in terms of a simple recursion relation. This formalism,
that is closely related to the Kodaira-Spencer formulation of the
B-model, can be viewed as the bosonized version of our fermionic
formulation. It would be interesting to understand this
non-commutative version of the familiar boson/fermion correspondence
and its interpretation in terms of $\cD$-modules in more detail.

\vspace{12mm}

\centerline{\bf Acknowledgments} 

We would like to thank M.~Aganagic, D.~Ben-Zvi, S.~Cherkis, B.~Eynard,
J.~Manschot, W.~Nahm, M.~Rocek, G.~Segal, E.~Verlinde, E.~Witten for
discussions.

We would like to thank the 5th Simons Workshop in Mathematics and Physics
at Stony Brook (C.V. and R.D.) and the Aspen Center for Physics (R.D.)
for inspiring surroundings and discussions. The research of R.D., L.H. and  
P.S. was supported by a NWO Spinoza
grant and the FOM program {\it String Theory and Quantum Gravity}. 
P.S. was also supported by MNiSW grant N202-004-31/0060.
The research of C.V. was supported in part by NSF grants PHY-0244821
and DMS-0244464. 
 
\newpage

\appendix

\section{Level-rank duality and $U(Nk)_1$ decomposition}  \label{app-decompose}

The affine algebras $\widehat{su}(N)_k$ and $\widehat{su}(k)_N$ are
related by the so-called level-rank duality
\cite{frenkel,jimbo-miwa,nakanishi,nrs,Hasegawa}, which maps to
each other orbits of their irreducible integrable representations under
outer automorphism groups. Let us explain this in more detail. The
Dynkin diagram of $\widehat{su}(N)_k$ consists of $N$ nodes permuted
in a cyclic order by the outer automorphism group $\mathbb{Z}_N$. This
induces also an action on affine irreducible integrable
representations. There are
\be
\frac{(N+k-1)!}{(N-1)!\,k!}   \label{nr-weights}
\ee
such representations of $\widehat{su}(N)_k$, which can be identified
in a standard way with Young diagrams $\rho$ with at most $N-1$ rows
and at most $k$ columns. We denote the set of such diagrams by
$\cY_{N-1,k}$. In particular, the generator of the outer automorphism
group $\sigma_N$, the so-called basic outer automorphism, has a simple
realization in terms of a Young diagram
$\rho=(\rho_1,\ldots,\rho_{N-1})$ corresponding to a given integrable
representation. The action of $\sigma_N$ amounts to adding a row of
length $k$ as a first row of $\rho$, and then reducing the diagram,
{\it i.e.} removing $\rho_{N-1}$ columns which acquired a length $N$ (so
that indeed $\sigma_N(\rho)\in \cY_{N-1,k}$),
\be 
\sigma_N (\rho_1,\ldots,\rho_{N-1}) =
(k-\rho_{N-1},\rho_1-\rho_{N-1},\ldots,\rho_{N-2}-\rho_{N-1}).  \label{cyclic}
\ee
It follows that $\sigma_N^N(\rho)=\rho$, as expected for
$\mathbb{Z}_N$ symmetry. All $N$ irreducible integrable
representations related by an action of $\sigma_N$ constitute an orbit
denoted as $[\rho] \subset \cY_{N-1,k}$. As an example, the  $\mathbb{Z}_4$
orbit generated from $\widehat{su}(4)_3$ irreducible integrable
representation corresponding to a diagram $\rho=(2,1)\in \cY_{3,3}$ is
given by
$$ 
\Yvcentermath1 \yng(2,1) \ \to \ \yng(3,2,1) \ \to \ \yng(2,2,1)
\ \to \ \yng(2,1,1)
$$

The number of such $\mathbb{Z}_N$ orbits is given by
(\ref{nr-weights}) divided by $N$. For both $\widehat{su}(N)_k$ and
$\widehat{su}(k)_N$ this number is the same, therefore a bijection
between orbits of their integrable irreducible representations
exists. The level-rank duality is a statement that for
$\widehat{su}(N)_k$ orbit represented by a diagram
$\rho\in\cY_{N-1,k}$ there is a canonical bijection realized as
\bea 
\cY_{N-1,k} \supset [\rho] & =&
\{\sigma^j_N(\rho)\ |\ j=0,\ldots,N-1 \}\ \mapsto \nonumber \\ 
& \mapsto & \{\sigma^a_k(\rho^t)\ |\ a=0,\ldots,k-1\} = [\rho^t]
\subset \cY_{k-1,N}, \label{lr-orbit} \eea
where $^t$ denotes a transposition and a diagram $\rho^t$ should be
reduced (i.e. all columns of length $k$ should be removed if $\rho_1$
was equal to $k$).

The level-rank duality can also be formulated in terms of the
embedding
$$ 
\widehat{u}(1)_{Nk} \times \widehat{su}(N)_k \times
\widehat{su}(k)_N \subset \widehat{u}(Nk)_1.
$$
The $\widehat{u}(Nk)_1$ affine Lie algebra can be realized in terms of
$Nk$ free fermions, so that their total Fock space
$\mathcal{F}^{\otimes Nk}$ decomposes under this embedding as
\be
\cF^{\otimes Nk} = \bigoplus_{\rho} U_{\|\rho\|} \otimes V_{\rho} \otimes
{\wt V}_{\wt\rho},  \label{lr-fock}
\ee
where $U_{\|\rho\|}$, $V_{\rho}$ and ${\wt V}_{\wt\rho}$ denote
irreducible integrable representations of $\widehat{u}(1)_{Nk}$,
$\widehat{su}(k)_N$, and $\widehat{su}(N)_k$ respectively. In the
above decomposition only those pairs $(\rho,\wt\rho)$ arise, which
represent orbits mapped to each other by the duality
(\ref{lr-orbit}). For a given $\widehat{su}(N)_k$ orbit $[\rho]$
represented by $\rho$, these pairs are therefore of the form
$(\sigma^j_N(\rho),\sigma^a_k(\rho^t))$, where $\sigma_N$ and
$\sigma_k$ are appropriate outer automorphism groups. The $U(1)$
charge corresponding to such a pair is $\|\rho\|=(|\rho|+jk +
aN)\ \textrm{mod}\ Nk$, where $|\rho|$ is the number of boxes in the
Young diagram $\rho$. With such identifications, the decomposition
(\ref{lr-fock}) can be written in terms of characters as
\cite{Hasegawa}
\be
\chi^{\widehat{u}(Nk)_1}(u,v,\tau) = \sum_{[\rho]\subset
  \mathcal{Y}_{N-1,k}} \sum_{j=0}^{N-1} \sum_{a=0}^{k-1}
\chi^{\widehat{u}(1)_{Nk}}_{|\rho|+jk+aN}(N |u| + k
|v|,\tau)\ \chi^{\widehat{su}(N)_k}_{\sigma^j_N(\rho)} (\overline{u},\tau)
\chi^{\widehat{su}(k)_N}_{\sigma^a_k(\rho^t)}
(\overline{v},\tau).   \label{uNk-decompose}
\ee
Here $u=(u_j)_{j=1...N}$ are elements of the Cartan subalgebra of $u(N)$,
$|u|=\sum_j u_j$ and $\overline{u}$ denotes the traceless
part. Similarly $v=(v_a)_{a=1...k}$ are elements of Cartan subalgebra
of $u(k)$.  $\chi^{\widehat{su}(N)_k}_{\rho} (\overline{u},\tau)$ are
characters of $\widehat{su}(N)_k$ at level $k$ for an integrable
irreducible representation specified by a Young diagram $\rho$, and
$\chi^{\widehat{u}(1)_N}_j$ characters are defined as
$$ 
\chi^{\widehat{u}(1)_N}_j(z,\tau) =
\frac{1}{\eta(q)}\sum_{n\in\mathbb{Z}} q^{\frac{N}{2}(n+j/N)^2}
e^{2\pi i z (n+j/N)}
$$
for $q=e^{2\pi i \tau}$. 

As an example of a decomposition (\ref{lr-fock}) let us consider the
case of $\widehat{u}(1)_{12} \times \widehat{su}(4)_3 \times
\widehat{su}(3)_4 \subset \widehat{u}(12)_1$, with $N=4$ and
$k=3$. From (\ref{nr-weights}) we deduce there are 5 orbits of outer
automorphism groups $\mathbb{Z}_4$ and $\mathbb{Z}_3$. Let us consider
$\widehat{su}(4)_3$ integrable representation related to a diagram
$\rho=\square$, and the corresponding $\widehat{su}(3)_4$ diagram
$\rho^t = \square$. The two orbits under $\sigma_4$ and $\sigma_3$ are
shown respectively in the first row and column of a table below. All
12 pairs of representations appear in the decomposition
(\ref{lr-fock}) with $\widehat{u}(1)_{12}$ charges given in the
table. Note that acting with $\sigma_4$ takes us to another pair of
weights given by a step to the right in the table, and increases
$\widehat{u}(1)_{12}$ charge by 3 (modulo 12). The action of
$\sigma_3$ takes us a step to the bottom in the table and increases
$\widehat{u}(1)_{12}$ charge by 4 (modulo 12). Of course the same
table would be generated if we started building it from any other
element of these two orbits.
\medskip

$$
\Yvcentermath1
\begin{array}{c|ccccccc}
   & \yng(1) & \to & \yng(3,1) & \to & \yng(3,3,1) & \to & \yng(2,2,2) \\ 
\\ \hline
\\
\yng(1) & 1 & & 4 & & 7 & & 10  \\ 
\downarrow & & & & & \\
\yng(4,1) & 5 & & 8 & & 11 & & 2 \\
\downarrow & & & & &\\
\yng(3,3) & 9 & & 0 & & 3 & & 6
\end{array}
$$
\medskip

Pairs of $\widehat{su}(4)_3 \times \widehat{su}(3)_4$ integrable
weights with the same fixed $\widehat{u}(1)_{12}$ charge arising in
the decomposition of $\widehat{u}(12)_1$ are easily found if all 5
such tables of orbits are drawn. For example for charge 0 we then get
$$
\Yvcentermath1
\bullet\otimes\bullet\ +\ \yng(3,1)\otimes\yng(3,3)\ +\ \yng(2,2)\otimes\yng(4,2)\ + 
$$
$$
\Yvcentermath1
+\ \yng(3,3,2)\otimes\yng(3)\ + \ \yng(2,1,1)\otimes\yng(2,1)
$$

\end{document}